\documentclass[useAMS,usenatbib]{mn2e}
\usepackage{graphicx}
\def\apj{ApJ}%
\def\apjl{ApJ}%
\def\apjs{ApJS}%
\def\aap{A\&A}%
\def\mnras{MNRAS}%
\title[The role of thermodynamics in disc fragmentation ]
{The role of thermodynamics in disc fragmentation}
\author[D. Stamatellos \& A.~P. Whitworth]
{Dimitris Stamatellos\thanks{E-mail:D.Stamatellos@astro.cf.ac.uk} and Anthony~P. Whitworth \\ 
School of Physics \& Astronomy,Cardiff University, Cardiff, CF24 3AA, Wales, UK}

\begin{document}

\date{Accepted 2009 . Received 2009 May 24; in original form 2009 May 24}

\pagerange{\pageref{firstpage}--\pageref{lastpage}} \pubyear{2009}

\maketitle

\label{firstpage}

\begin{abstract}
Thermodynamics play an important role in determining the way a  protostellar disc  fragments  to form planets, brown dwarfs and low-mass stars. We explore the effect that different treatments of radiative transfer have in simulations of fragmenting discs. Three prescriptions for the radiative transfer are used, (i) the diffusion approximation of Stamatellos et al. (2007a), (ii) the barotropic equation of state (EOS) of Goodwin et al. (2004a), and (iii) the barotropic EOS of Bate et al. (2003).  The barotropic approximations capture the general evolution of the density and temperature at the centre of each proto-fragment  but (i) they do not make any adjustments the particular circumstances of a proto-fragment forming in the disc, and (ii) they do not take into account thermal inertia effects that are important for fast-forming proto-fragments in the outer disc region.  As a result, the number of fragments formed in the disc and their properties  are different,  when a barotropic EOS is used.  This is important not only for disc studies but also for simulations of collapsing turbulent clouds, as in many cases in such simulations stars form with discs that subsequently fragment. 
We also examine the difference in the way proto-fragments condense out in the disc at different distances from the central star using the diffusion approximation  and following the collapse of each proto-fragment until the formation of the second core ($\rho\simeq10^{-3}~{\rm g\ cm}^{-3}$). We find that proto-fragments forming closer to the central star tend to form earlier and evolve faster from the first  to the second core than proto-fragments forming in the outer disc region. The former have a large pool of material in the inner disc region that they can accrete from and grow in mass. The latter accrete more slowly and they are hotter because they generally form  in a quick abrupt event.

\end{abstract}


\section{Introduction}

The formation of discs around young stars is an integral part of the star formation process. The presence of discs has long been inferred from the spectral energy distributions of young Pre-Main Sequence stars; these show long-wavelength emission in excess of the photospheric emission from the stellar surface, and this long-wavelength emission  is attributed to a  disc.  Moreover,  in the last 15 years, discs have been imaged directly, both in absorption and emission with the Hubble Space Telescope and with ground bases telescopes using adaptive optics (e.g. Watson et al. 2007).
 
On the theoretical front discs have been studied for over 30 years (e.g. Lynden-Bell \& Pringle 1974). Discs around young stars are a natural outcome of the angular momentum of the parental cores that collapse to form these stars. Most of the matter from such a core cannot accrete directly onto the young protostar but firstly falls onto the disc and then spirals inwards onto the protostar. The disc matter has to redistribute its angular momentum so that it can accrete onto the protostar. An important role in this process is played by the effective disc viscosity (e.g. Clarke 2009). The origin of the disc viscosity is still undetermined. It is believed that gravitational instabilities (Toomre 1964; Lin \& Pringle 1987) or magneto-rotational instabilities (Balbus \& Hawley 1991)  may play an important role in redistributing angular momentum in discs. 

Thermodynamics play an important role in determining whether a disc can fragment to form planets, brown dwarfs and low-mass stars. For fragmentation to occur the disc must satisfy the Toomre criterion (Toomre 1964), i.e. it has to be massive enough so that gravity can overcome thermal and local centrifugal support, 
\begin{equation}
Q(R)\equiv\frac{c(R)\,\kappa(R)}{\pi\,G\,\Sigma(R)}\;\la\;1\,;
\end{equation}
here $Q$ is the Toomre parameter, $c$ is the sound speed, $\kappa$ is the epicyclic frequency, and $\Sigma$ is the surface density.  Gammie  (2001) has pointed out that  for a disc to fragment, the disc must also cool efficiently so that proto-condensations forming in the disc do not simply undergo an adiabatic bounce and dissolve. Both theory and simulations (Gammie 2001; Johnson \& Gammie 2003; Rice et al. 2003, 2005;  Mayer et al. 2004; Mejia et al. 2005; Stamatellos et al. 2007b) indicate that for disc fragmentation to occur, the cooling time must satisfy
\begin{equation}
t_{_{\rm COOL}}\;\,<\;\;{\cal C}(\gamma)\,t_{_{\rm ORB}}\,,\hspace{1.0cm}0.5\la{\cal C}(\gamma)\la 2.0\,;
\end{equation}
here $t_{_{\rm ORB}}$ is the local orbital period, and $\gamma$  is the adiabatic exponent. 
The factor ${\cal C}(\gamma)$ is found to be higher for lower $\gamma$ (e.g. higher for  $\gamma=7/5$ than for $\gamma=5/3$; Rice et al. 2005). It also depends on how fast the cooling rate reduces due to secular disc evolution from an initial higher value (Clarke et al. 2007).

Limitations in computing power have made it impossible to study the evolution of discs including a full treatment  for the radiative transfer (i.e. multi-frequencies, 3 dimensions). Traditionally the effect of radiative transfer in hydrodynamic simulations of star formation has been taken into account using a barotropic EOS, e.g.
\begin{equation}
\label{eq:beos}
\frac{P(\rho)}{\rho}\equiv c_s^2(\rho)=c_0^2\ \left[1+\left(\frac{\rho}{\rho_{\rm crit}}\right)^{\gamma-1}\right],
\end{equation}
where P is the pressure, $\rho$ is the density, $c_s$ is the isothermal sound speed and $c_0$ is the isothermal sound speed in the low-density gas (e.g. $c_0=0.19$ km s$^{-1}$ for molecular hydrogen at 10 K). The use of a barotropic EOS aims to reproduce the evolution of the density and temperature at the centre of a collapsing fragment. Initially at densities $<10^{-13}$g cm$^{-3}$ the gas is approximately isothermal, hence $P\propto\rho$. When the density increases above the critical value $\rho_{\rm crit}$ $\sim 3\times 10^{-13}$g cm$^{-3}$, the center of the fragment becomes optically thick and the collapse proceeds adiabatically, hence $P\propto\rho^\gamma$.

The adiabatic exponent is initially 5/3 (molecular hydrogen behaves effectively as a monatomic gas) and then becomes 7/5 when the hydrogen molecules are rotationally excited. Usually $\rho_{\rm crit}$ and $\gamma$ are set so as to match the evolution at the centre of a 1-M$_{\sun}$ molecular cloud (e.g. Larson 1969; Masunaga \& Inutsuka 2000). This is not ideal as (i) it does not account of the fact that the evolution of temperature and density away from the centre of the cloud is not the same as in the cloud centre, (ii) it ignores the fact that lower mass fragments become optically thick at larger densities than higher mass fragments, and (iii) it ignores the effects of thermal inertia, i.e. situations where the thermal timescale is comparable with, or longer than, the dynamical timescale (Boss et al. 2000). Previous studies have shown that the outcome of cloud fragmentation is sensitive to the values of $\rho_{\rm crit}$ and $\gamma$ used (e.g. Li et al. 2003; Jappsen et al. 2005; Larson 2005).

In the last few years the increase in computing power has made possible to use the diffusion approximation to treat the effects of radiative transfer in high-resolution simulations of low- and high-mass star formation (Boss et al. 2000; Whitehouse \& Bate 2006; Stamatellos et al. 2007a; Krumholz et al. 2007; Mayer et al. 2007; Forgan et al. 2009). 

In this paper we examine the effect of  different treatments of the energy equation on hydrodynamic simulations of disc fragmentation. We focus on discs that are massive enough for the Toomre criterion to be satisfied and extended enough for the Gammie criterion to be satisfied in their outer regions. Such discs will inevitably  fragment and we seek to determine how the number of fragments and their properties (e.g. mass, radius of formation) depend on the disc thermodynamics. We also want to determine any differences in the properties of fragments that are produced at different radii in the disc. This is important not only for disc studies but also for   simulations of collapsing turbulent clouds as in many cases in such simulations stars form with a disc that eventually fragments to spawn a second generation of low-mass stars and brown dwarfs.

The effects of radiative transfer are treated in three ways, (i) with the  diffusion approximation of Stamatellos et al. (2007a), (ii) with the barotropic EOS of Goodwin et al. (2004a), and (iii) with the barotropic EOS of Bate et al. (2003). We find that the characteristics of disc fragmentation are different in  these three treatments. The number of fragments condensing out in the disc are fewer for the Goodwin et al. barotropic EOS than for the Stamatellos et al. approximation or the Bate et al. (2003) barotropic EOS.  This is expected as the Goodwin et al. EOS is harder, i.e. matter is hotter at a given density than in the other two treatments. We also find that there are differences in the evolution of density and temperature in the fragments that form in the disc between the barotropic approximations and the Stamatellos et al. (2007a) diffusion approximation. Hence, the properties of the objects formed by fragmentation (e.g. number of objects, masses, binarity) in simulations using a barotropic EOS (e.g. Bate et al. 2003; Goodwin et al. 2004a; Bate et al. 2009a) are to  be treated with caution (cf. Attwood et al. 2009).

The paper is structured as follows. In Section 2, we  describe the initial conditions of the discs we examine. In Section 3, we describe the numerical method we use to perform the simulations and the three treatments of the energy equation used. In Section 4 we discuss the differences between the three treatments, and in Section 5 we investigate differences between fragments forming at different disc radii, when the diffusion approximation is used.  In Section 6 we discuss the resolution achieved by our simulations. Finally, in Section 7 we summarize our results and discuss their implications for simulations of star formation.

\section{Disc initial conditions}

We shall assume a star-disc system in which the central primary star  has initial mass $M=0.7\,{\rm M}_{\sun}$. Initially the disc has mass $M_{_{\rm D}}=0.7\,{\rm M}_{\sun}$, inner radius $R_{_{\rm IN}}=40\,{\rm AU}$, outer radius $R_{_{\rm OUT}}=400\,{\rm AU}$, surface density 
\begin{equation}
\Sigma_{_0}(R)=\frac{0.014\,{\rm M}_{\sun}}{{\rm AU}^2}\,\left(\frac{R}{\rm AU}\right)^{-7/4}\,,
\end{equation}
temperature
\begin{equation}\label{EQN:TBG}
T_{_0}(R)=250\,{\rm K}\,\left(\frac{R}{\rm AU}\right)^{-1/2}+10~{\rm  K}\,,
\end{equation}
and hence approximately uniform initial Toomre parameter $Q\sim 0.9$. Thus the disc is at the outset marginally gravitationally unstable at all radii. Such a disc can cool fast enough at radii $\stackrel{>}{_\sim} 70 AU$, hence it will fragment to spawn a number of objects having masses  from a few to a few hundred Jupiter masses (Whitworth et al. 2007; Stamatellos et al. 2007b; Stamatellos \& Whitworth 2009a,b; Boley 2009).  Observations suggest that the used density and temperature profiles are not unrealistic. Andrews et al. (2009) estimate that the disc surface density drops as $r^{-(0.4-1.0)}$, and  Osterloh \& Beckwith (1995) find that the disc temperature drops as $r^{-(0.35-0.8)}$.

Only a few extended massive discs have been observed around young protostars (e.g. Eisner et al. 2005, 2008; Rodriguez et al. 2005; Eisner \& Carpenter 2006). More recently, Andrews et al. (2009) reported discs in Ophiuchus with masses up to 0.14 M$_{\sun}$ and radii up to 200 AU. Stamatellos \& Whitworth (2009) argue that  such massive, extended discs may be more common during the initial stages of the formation of a protostar, but they rapidly dissipate (within a few thousand years) by gravitational fragmentation. For example the disc around the 0.33 M$_{\sun}$ protostar HL Tau has a mass of around 0.1 M$_{\sun}$ and extends to at least 100 AU, with a possible low-mass fragment at $\sim 65$~AU (Greaves et a. 2008). 

The formation of such discs is inevitable considering the angular momentum content of their parental cores. For example, a $1.4\,{\rm M}_\odot$ prestellar core with ratio of rotational to gravitational energy $\beta\equiv{\cal R}/|\Omega|$ will -- if it collapses monolithically -- forms a protostellar disc with outer radius $R_{_{\rm DISC}}\sim 400\,{\rm AU}\,(\beta/0.01)$. Since the observations of Goodman et al. (1993) indicate that many prestellar cores have $\beta \sim 0.02$, the formation of extended discs should be rather common. 

\section{Numerical method}

\subsection{Hydrodynamics}

We use the SPH code {\sc dragon} (Goodwin et al. 2004a,b; 2006), which invokes an octal tree (to compute gravity and find neighbours), adaptive smoothing lengths, multiple particle timesteps, and a second-order Runge-Kutta integration scheme. The code uses time-dependent artificial viscosity (Morris \& Monaghan 1996) with parameters $\alpha^\star=0.1$, $\beta=2\alpha$ and a Balsara switch (Balsara 1995), so as to reduce artificial shear viscosity (Artymowicz \& Lubow 1994; Lodato \& Rice 2004; Rice, Lodato, \& Armitage 2005). The number of neighbours is set to $N_{_{\rm NEIGH}}=50\pm0$ (Attwood et al. 2007). 

\subsection{Radiative transfer}
The effects of radiative transfer are treated in three ways, (i) with the  diffusion approximation of Stamatellos et al. (2007a), (ii) with the barotropic EOS of Goodwin et al. (2004a) (Eq.~\ref{eq:beos} with $\rho_{\rm crit}=10^{-13}$g cm$^{-3}$ and $\gamma=5/3$), and, (iii)
with the barotropic EOS of Bate et al. (2003)  (Eq.~\ref{eq:beos} with $\rho_{\rm crit}= 10^{-13}$g cm$^{-3} $ and $\gamma=7/5$).

\subsection{The Stamatellos et al. (2007a) diffusion approximation method}

The Stamatellos et al. (2007a) method uses the density and the gravitational potential of each SPH particle to estimate an optical depth $\tau_i$ for each particle through which the particle cools and heats. The net radiative heating rate for the particle is then
\begin{equation} 
\label{eq:radcool}
\left. \frac{du_i}{dt} \right|_{_{\rm RAD}} =
\frac{\, 4\,\sigma_{_{\rm SB}}\, (T_{_{\rm BGR}}^4-T_i^4)}{\bar{{\Sigma}}_i^2\,\bar{\kappa}_{_{\rm R}}(\rho_i,T_i)+{\kappa_{_{\rm P}}}^{-1}(\rho_i,T_i)}\,;
\end{equation}
here, the positive term on the right hand side represents heating by the background radiation field, and ensures that the gas and dust cannot cool radiatively below the background radiation temperature $T_{_{\rm BGR}}$. $\;{\kappa_{_{\rm P}}}(\rho_i,T_i)$ is the Planck-mean opacity,  $\sigma_{_{\rm SB}}$ the Stefan-Boltzmann constant, $\bar{{\Sigma}}_i$ is the mass-weighted mean column-density, and  $\bar{\kappa}_{_{\rm R}}(\rho_i,T_i)$, is the Rosseland-mean opacity (see Stamatellos et al. 2007a for details).

The method takes into account compressional heating, viscous heating, radiative heating by the background, and radiative cooling. It performs well, in both the optically thin and optically thick regimes, and has been extensively tested (Stamatellos et al. 2007a). In particular it reproduces the detailed 3D results of Masunaga \& Inutsuka (2000), Boss \& Bodenheimer (1979), Boss \& Myhill (1992), Whitehouse \& Bate (2006),  and also the analytic results of Spiegel (1957) and Hubeny (1990)

The gas is assumed to be a mixture of hydrogen and helium. We use an EOS (Black \& Bodenheimer 1975; Masunaga et al. 1998; Boley et al. 2007) that accounts (i) for the rotational and vibrational degrees of freedom of molecular hydrogen, and (ii) for the different chemical states of hydrogen and helium. We assume that ortho- and para-hydrogen are in equilibrium.

For the dust and gas opacity we use the parameterization  by Bell \& Lin (1994), $\kappa(\rho,T)=\kappa_0\ \rho^a\ T^b\,$, where $\kappa_0$, $a$, $b$ are constants that depend on the species and the physical processes contributing to the opacity at each $\rho$ and $T$. The opacity changes due to ice mantle melting, the sublimation of dust, molecular and H$^-$ contributions,  are all taken into account. 

\subsection{Sinks}

Sinks are used in the simulations in order to avoid very small timesteps when the density becomes very high. Sinks are created when a particle reaches a critical density and the particles in its neighbourhood (i.e. its 50 neighbours) are bound (Goodwin et al. 2004a,b; 2006).

The critical density for sink creation is different for the 3 treatments; (i) for the  diffusion approximation $\rho_{\rm sink}=10^{-3}$g cm$^{-3}$, (ii) for the Goodwin et al. barotropic EOS $\rho_{\rm sink}=5\times10^{-11}$g cm$^{-3}$, and (iii) for the Bate et al. barotropic EOS $\rho_{\rm sink}=10^{-13}$g cm$^{-9}$.  The reason for using  3 different sink creation threshold densities  is connected to the characteristics of each treatment. The  diffusion approximation of Stamatellos et al. (2007a) can follow self-consistently the evolution of a fragment to any given density. The only constraint is computational. For the Goodwin et al.  barotropic EOS a fragment becomes too hot as the density increases, thus no fragment would form if the sink creation density were set too high. The same holds for the Bate et al. barotropic EOS, but for higher densities (see Fig.~\ref{fig:td}).

The sink radius is set to 1~AU in all three treatments. Once a particle is within this radius and bound to the sink then it is accreted onto the sink and its mass is added to the mass of the sink.

\subsection{Radiative feedback}

Radiative feedback is included in the treatment using the  diffusion approximation. The radiation of the central star is taken into account by invoking a background radiation field with a temperature $T_{_{\rm BGR}}=T_{_0}(R)$ that is a function of the distance R (on the disc midplane) from this star. In effect, this means that, if the material in the disc is heated by compression and/or viscous dissipation, it can only cool radiatively if it is warmer than $T_{_0}(R)$ given by Eqn. (\ref{EQN:TBG}). 

Radiative feedback from brown dwarfs and stars formed in the disc is partially taken into account. Until a proto-fragment is replaced by a sink the effects of energy dissipation due to accretion of material onto the first and second core are treated self-consistently. Once a proto-fragment is replaced by a sink ($10^{-3}$~g~cm$^{-3}$) then  radiative feedback is ignored. We expect that because of this high sink creation density all of the fragments that form in our simulations are real; any possible additional fragmentation in the vicinity of each proto-fragment (within 10-20 AU) is  suppressed due to the heating of this region by energy that diffuses outwards  from the accretion shock around the proto-fragments.

Any additional heating due to accretion  material onto a sink is ignored as are deuterium and hydrogen burning (cf. Krumholz et al. 2007; Offner et al. 2009). The accretion luminosity could be important during the initial stages of the formation of an object when the accretion rate is high. The effect of nuclear burning is probably minimal. 

%%%%%%%%%%%%%%%%%%%%%%%%%%%%%%%%%%%%%%%%%%%%%
\begin{figure*}
\centering{
\includegraphics[height=5.7cm,angle=-90]{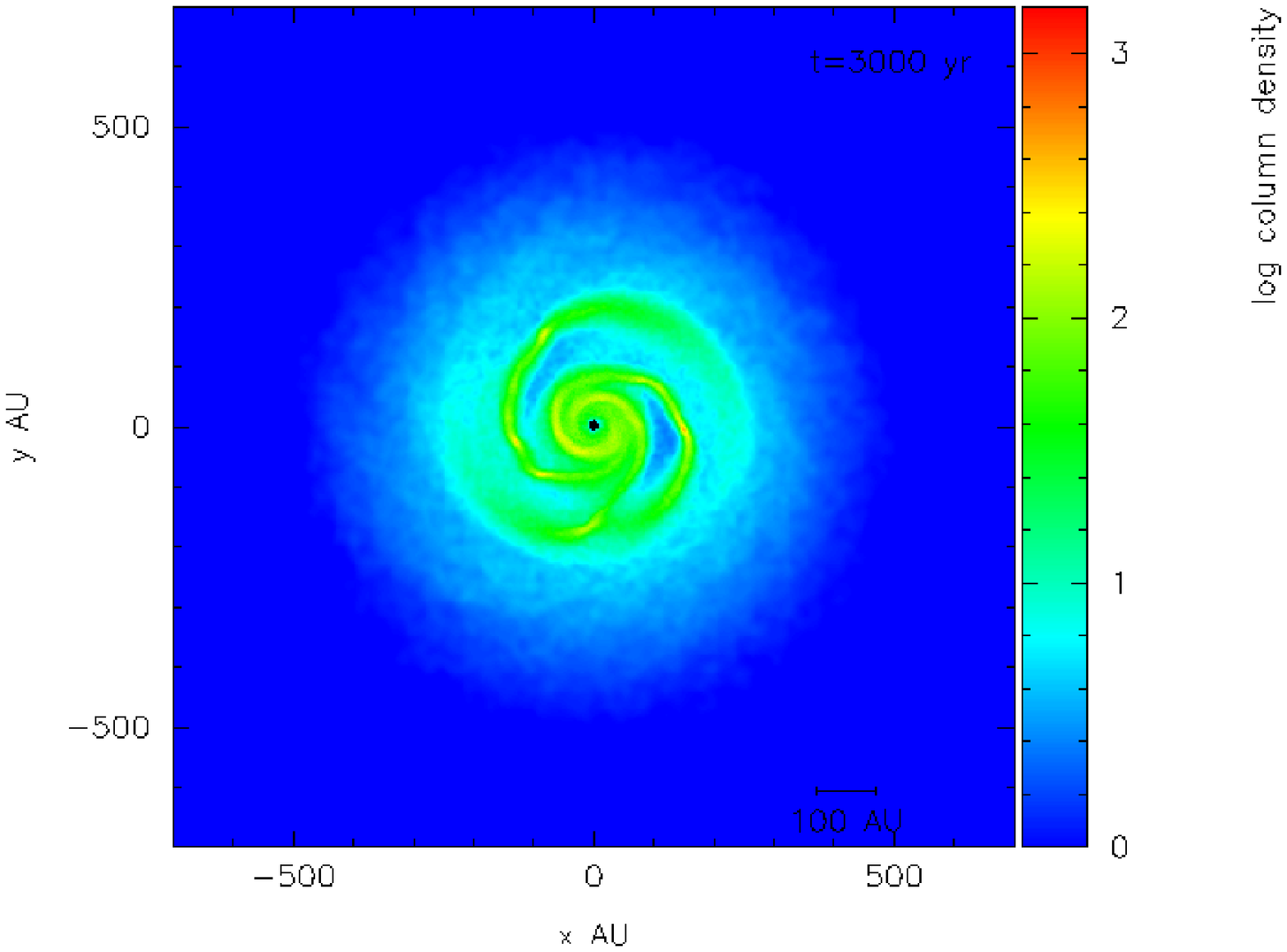}
\includegraphics[height=5.7cm,angle=-90]{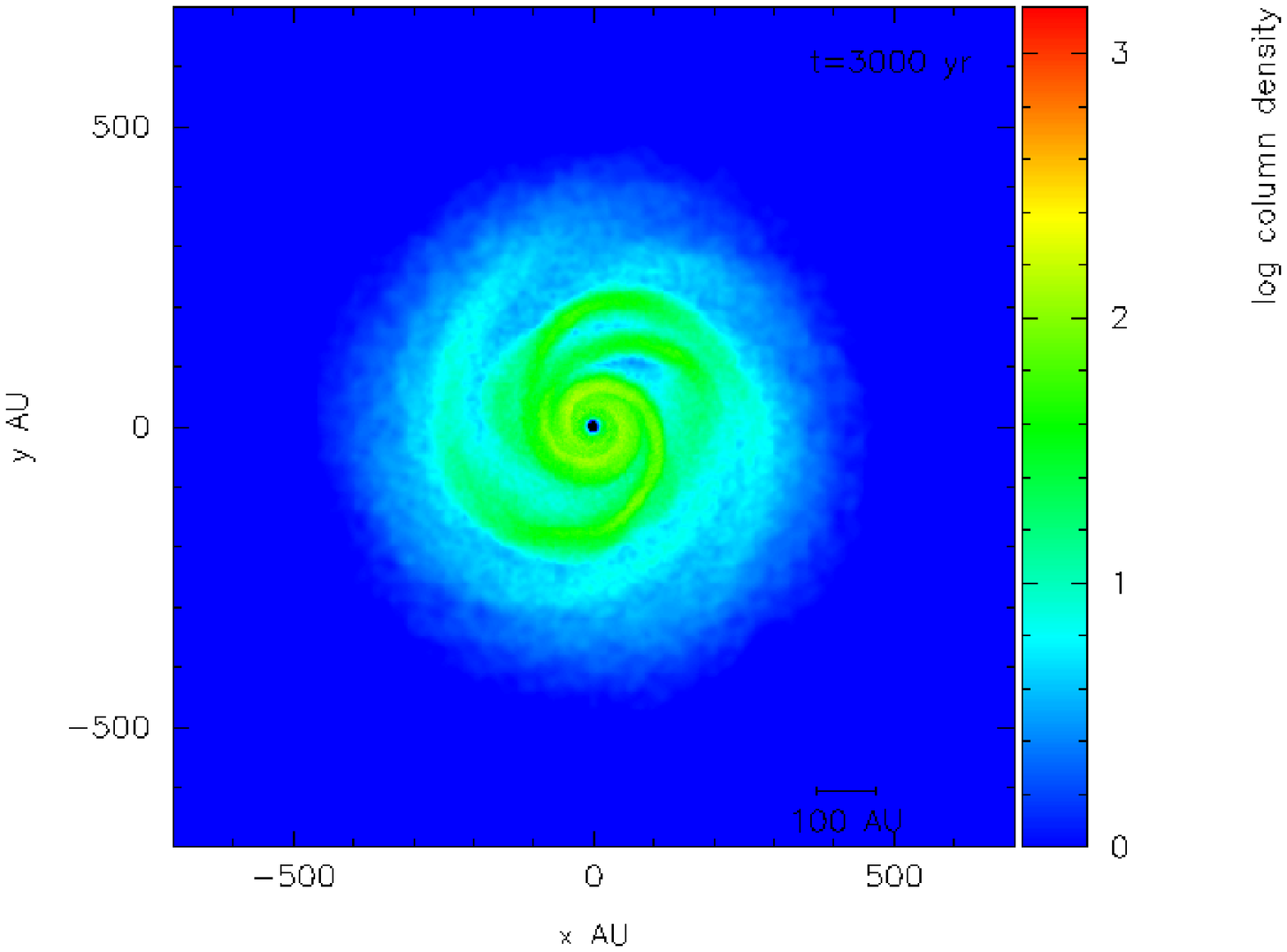}
\includegraphics[height=5.7cm,angle=-90]{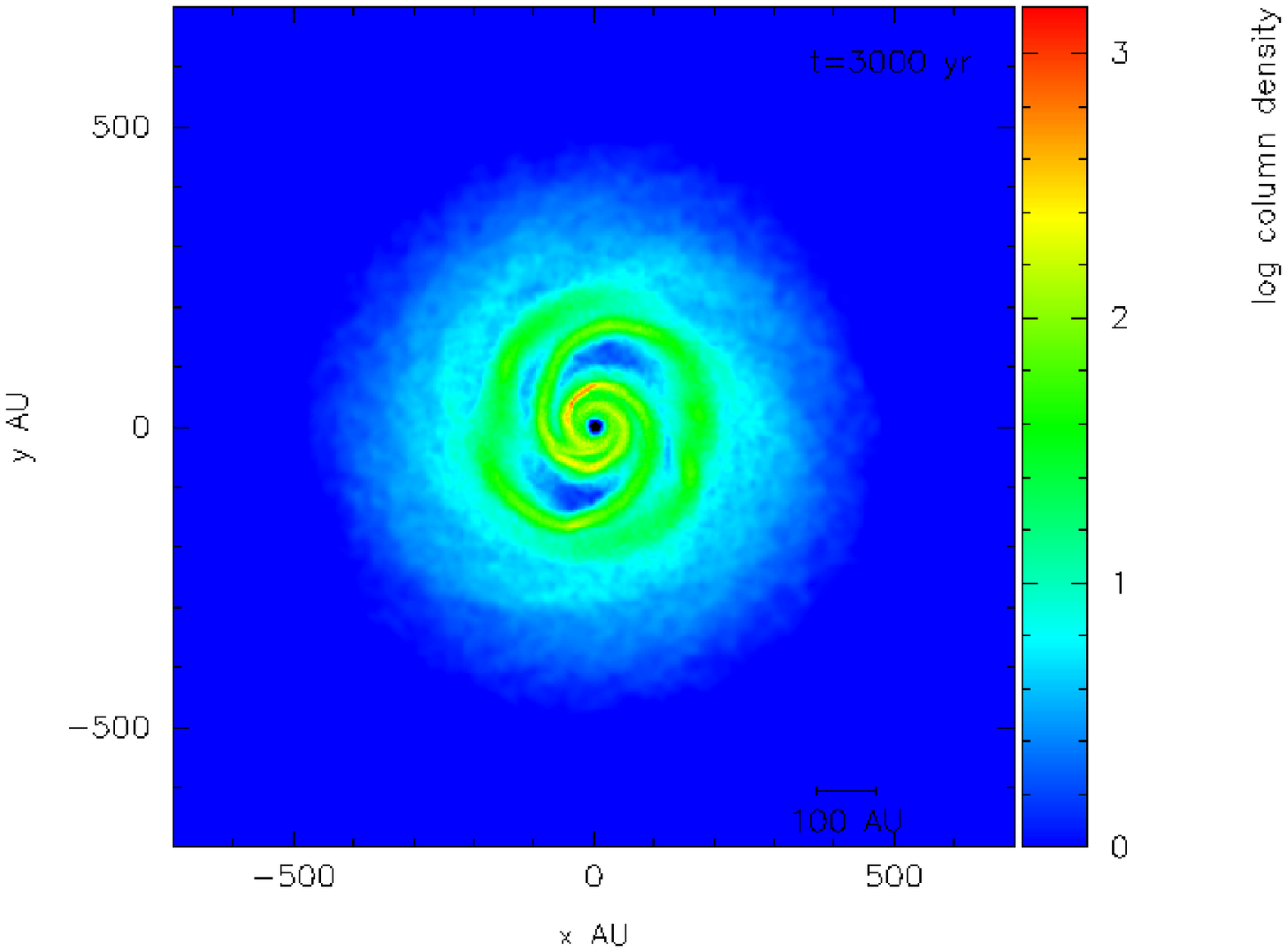}
}
\centering{
\includegraphics[height=5.7cm,angle=-90]{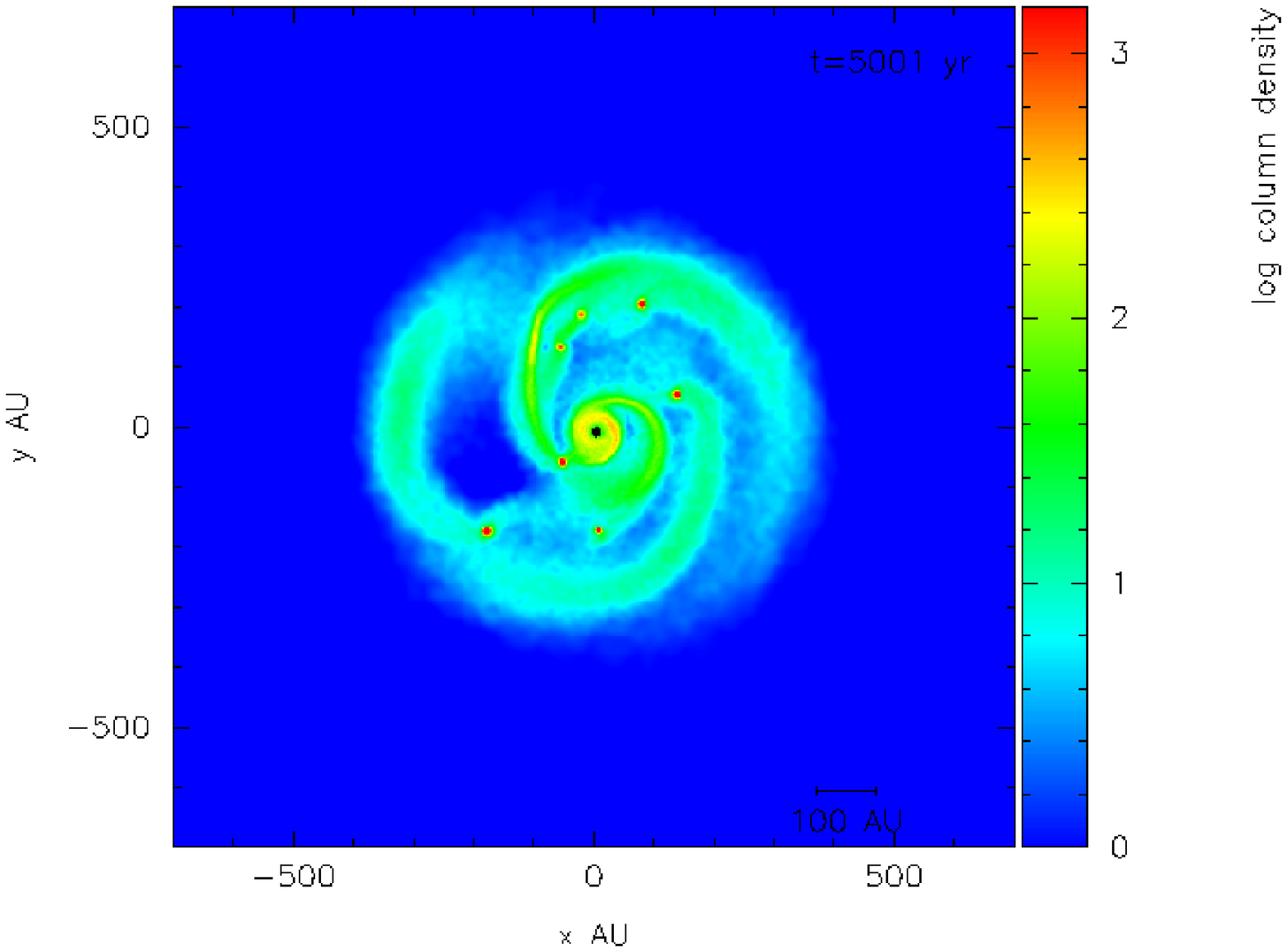}
\includegraphics[height=5.7cm,angle=-90]{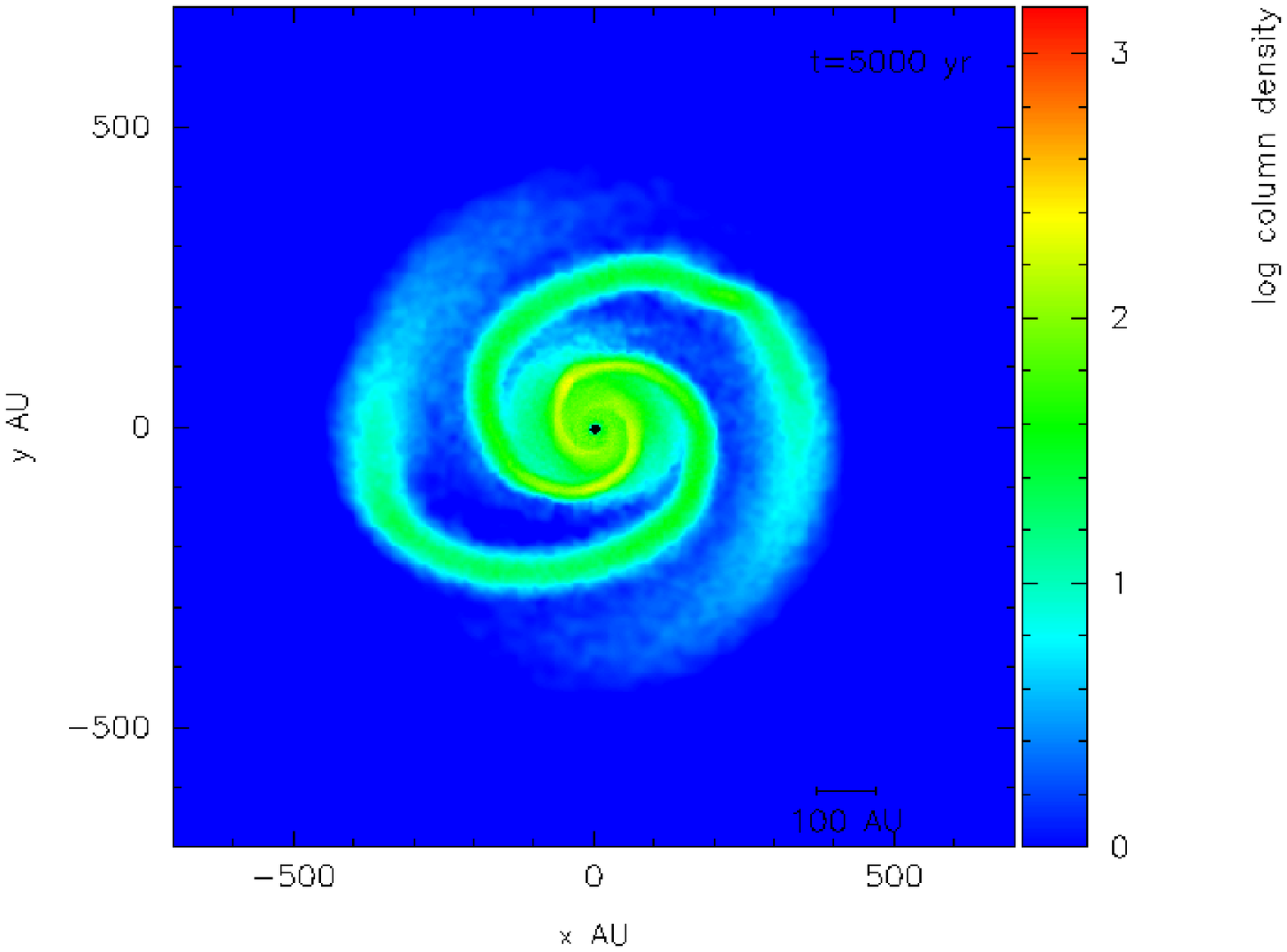}
\includegraphics[height=5.7cm,angle=-90]{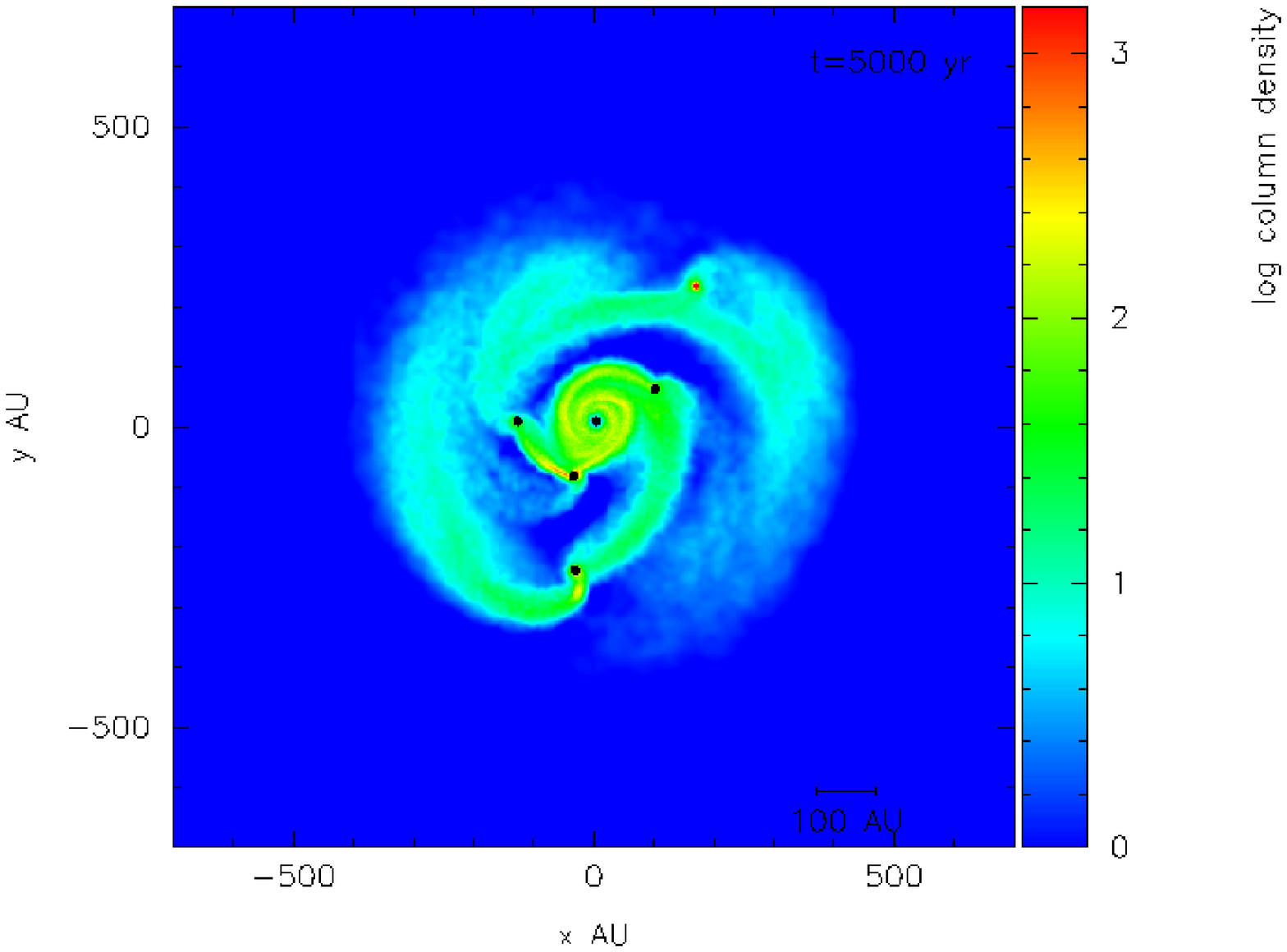}
}
\centering{
\includegraphics[height=5.7cm,angle=-90]{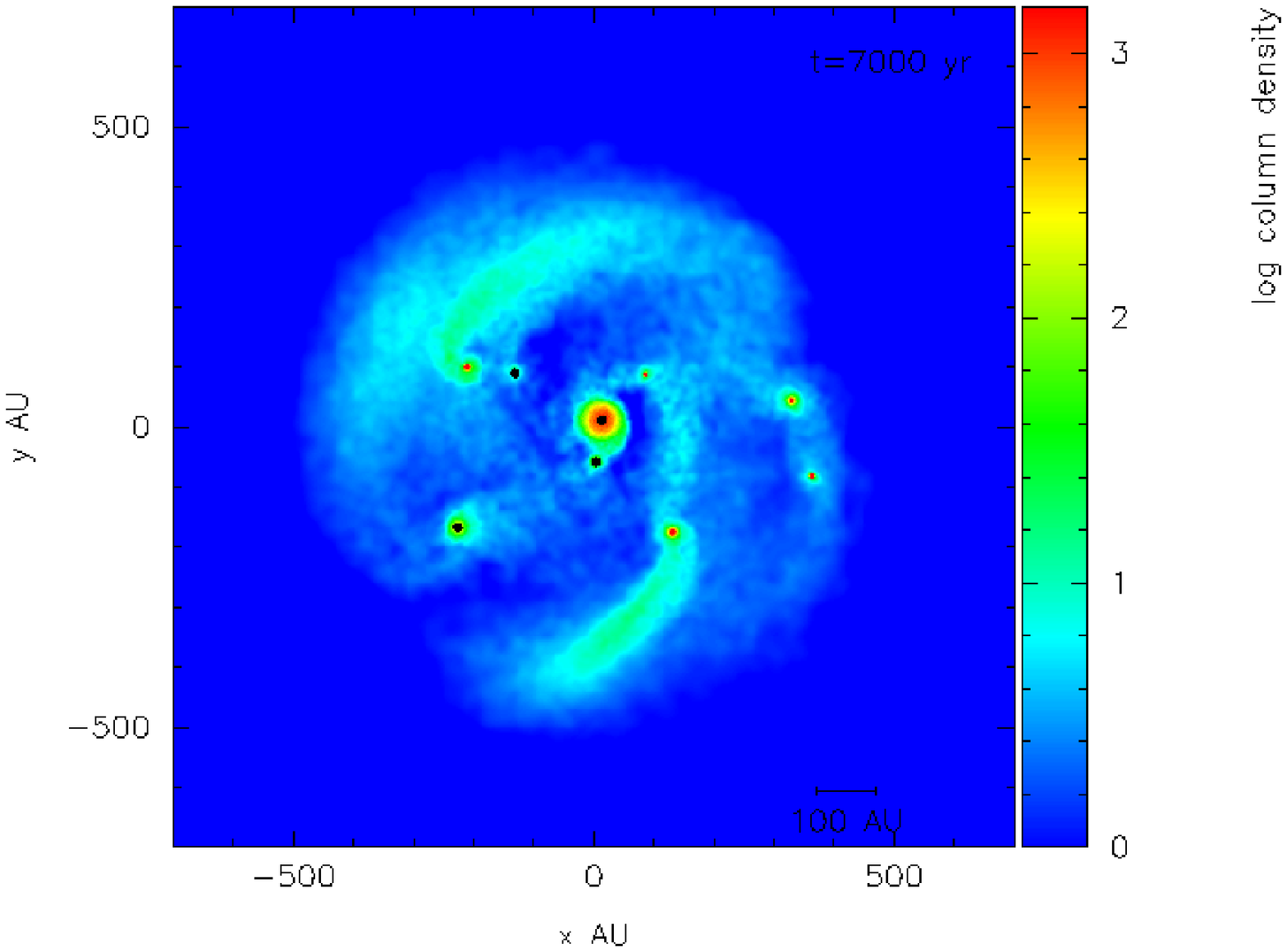}
\includegraphics[height=5.7cm,angle=-90]{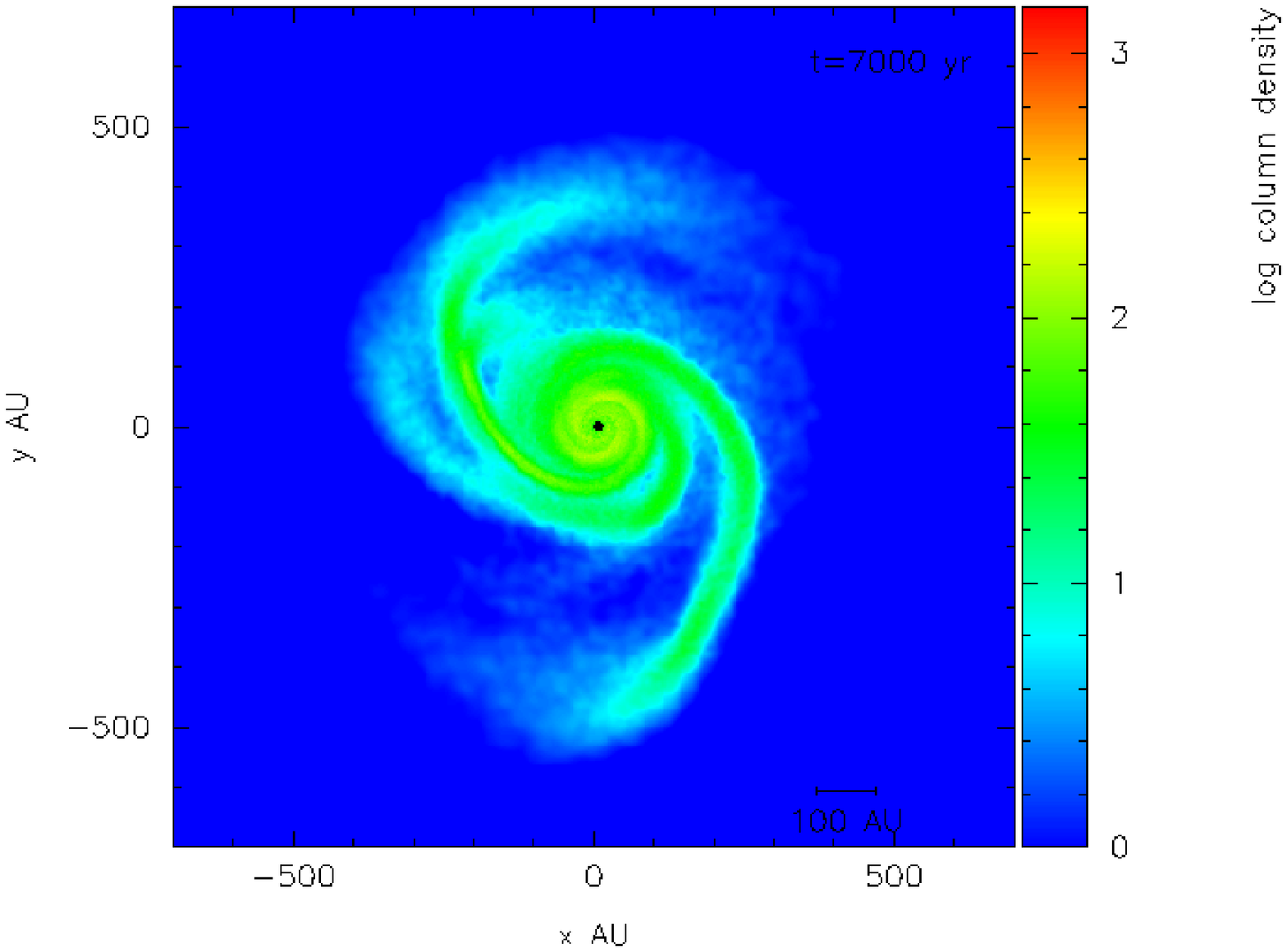}
\includegraphics[height=5.7cm,angle=-90]{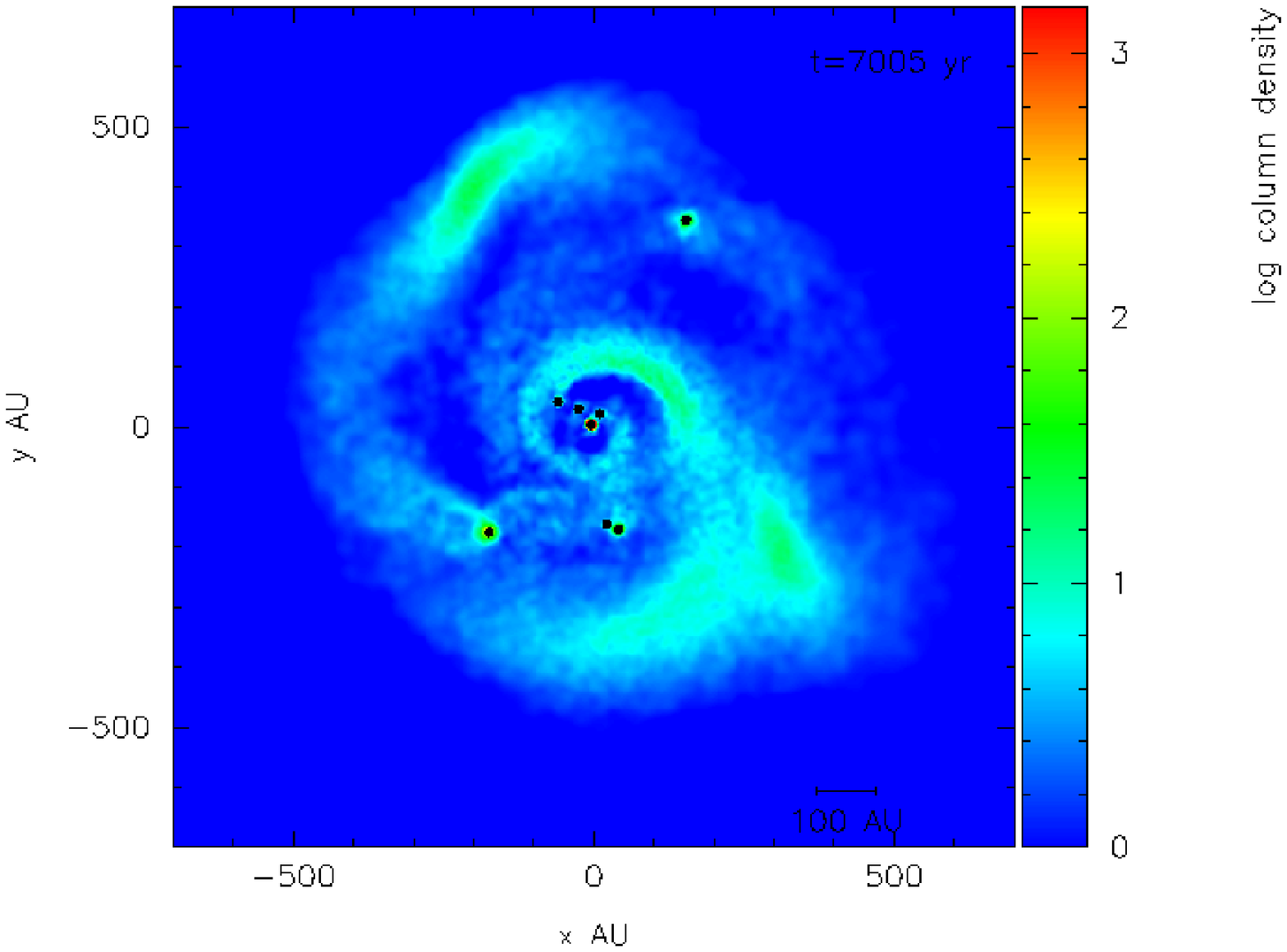}
}
\centering{
\includegraphics[height=5.7cm,angle=-90]{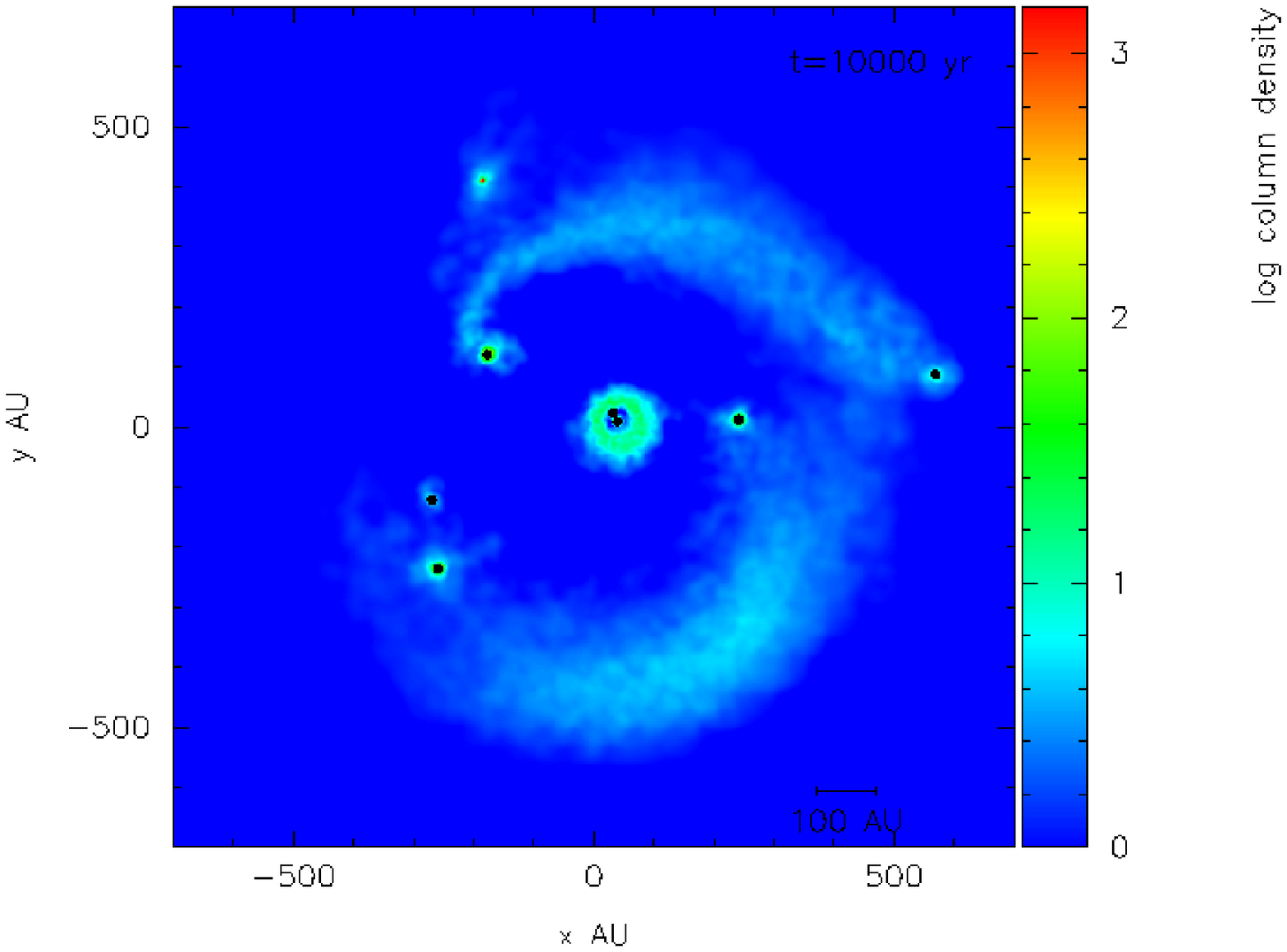}
\includegraphics[height=5.7cm,angle=-90]{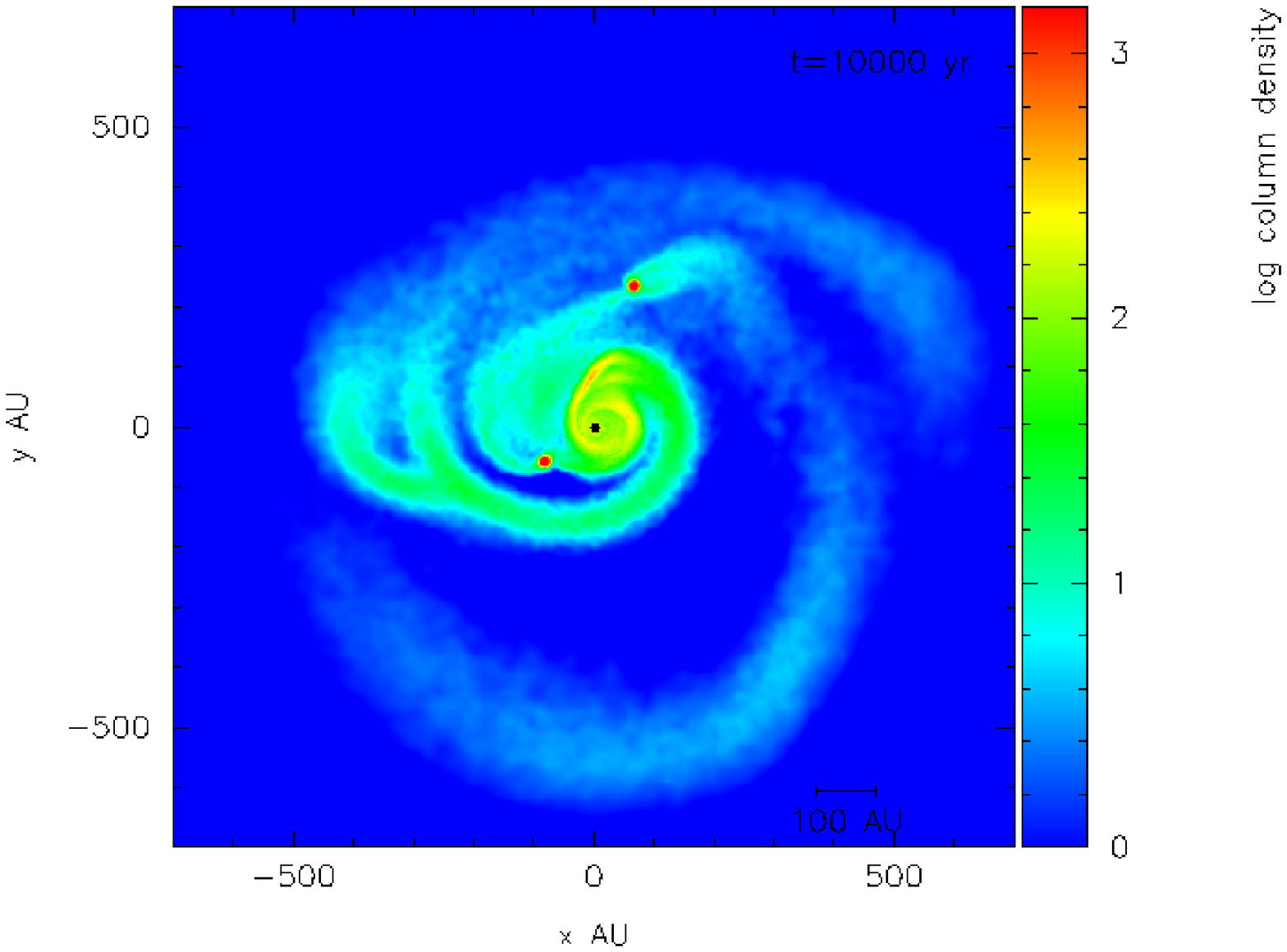}
\includegraphics[height=5.7cm,angle=-90]{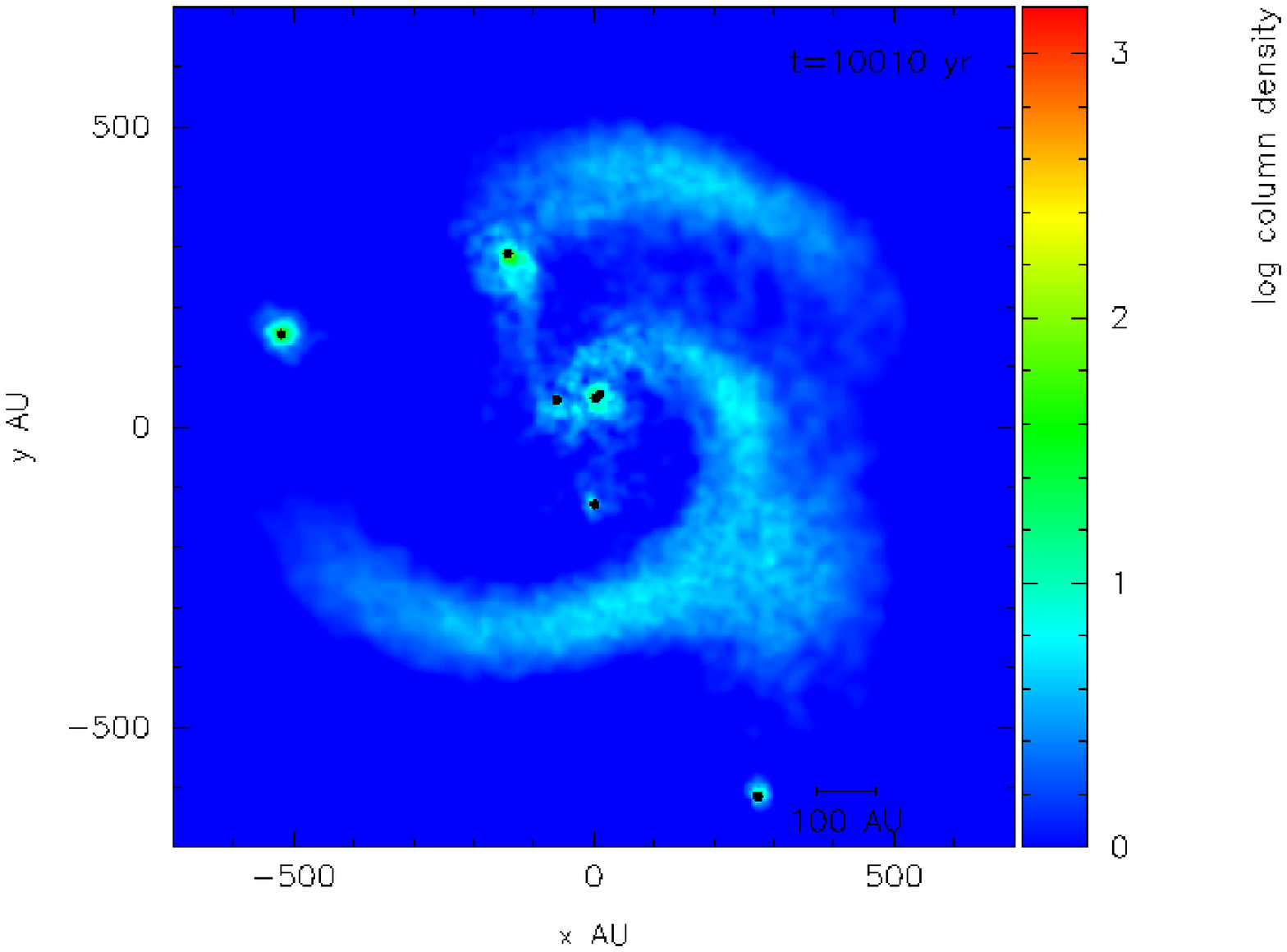}
}
\centering{
\includegraphics[height=5.7cm,angle=-90]{rt.15000.ps}
\includegraphics[height=5.7cm,angle=-90]{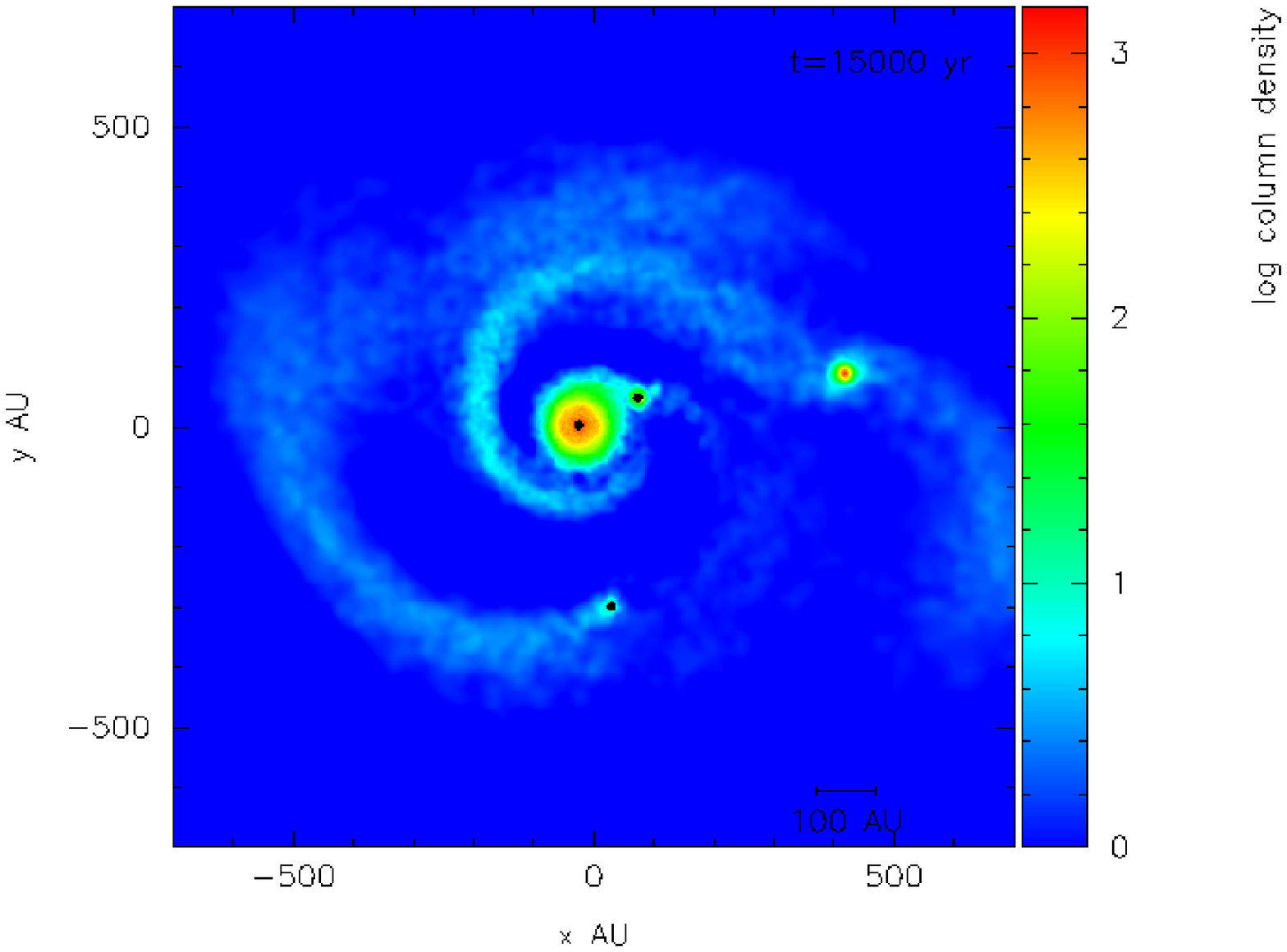}
\includegraphics[height=5.7cm,angle=-90]{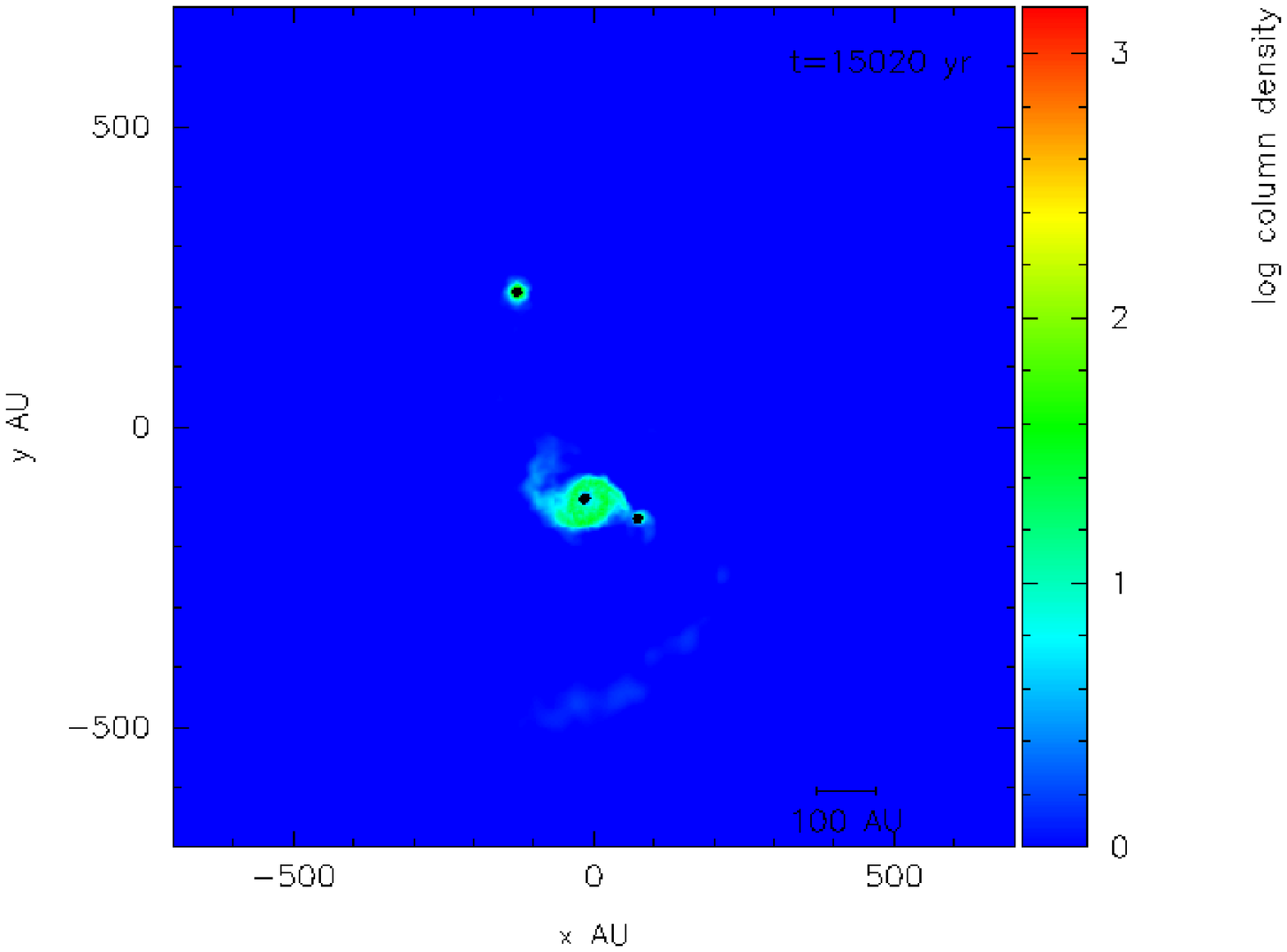}
}
\caption{Snapshots of the disc evolution (at times as marked on each column density plot), using (i) the Stamatellos et al. (2007a) diffusion approximation (left column) (ii) the Goodwin et al. (2004a) barotropic EOS (central column), and (iii) the Bate et al. (2003) barotropic EOS (right column). }
\label{fig:images}
\end{figure*}

\section{The  diffusion method vs the barotropic equation of state}

\begin{figure}
\centering{
\includegraphics[height=9cm,angle=-90]{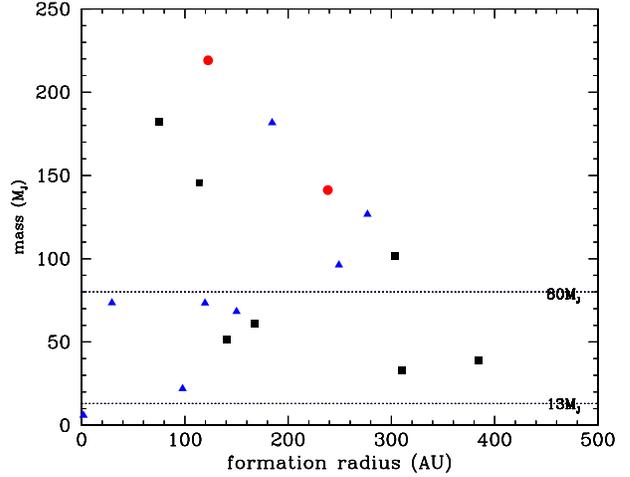}}
\caption{The mass of the objects formed by disc fragmentation and their formation radius, using (i) the Stamatellos et al. (2007a) diffusion approximation (squares) (ii) the Goodwin et al. (2004a) barotropic EOS (circles), and (iii) the Bate et al. (2003) barotropic EOS (triangles).}
\label{fig:mrsink}
\end{figure}

\begin{figure}
\centering{
\includegraphics[height=6.8cm]{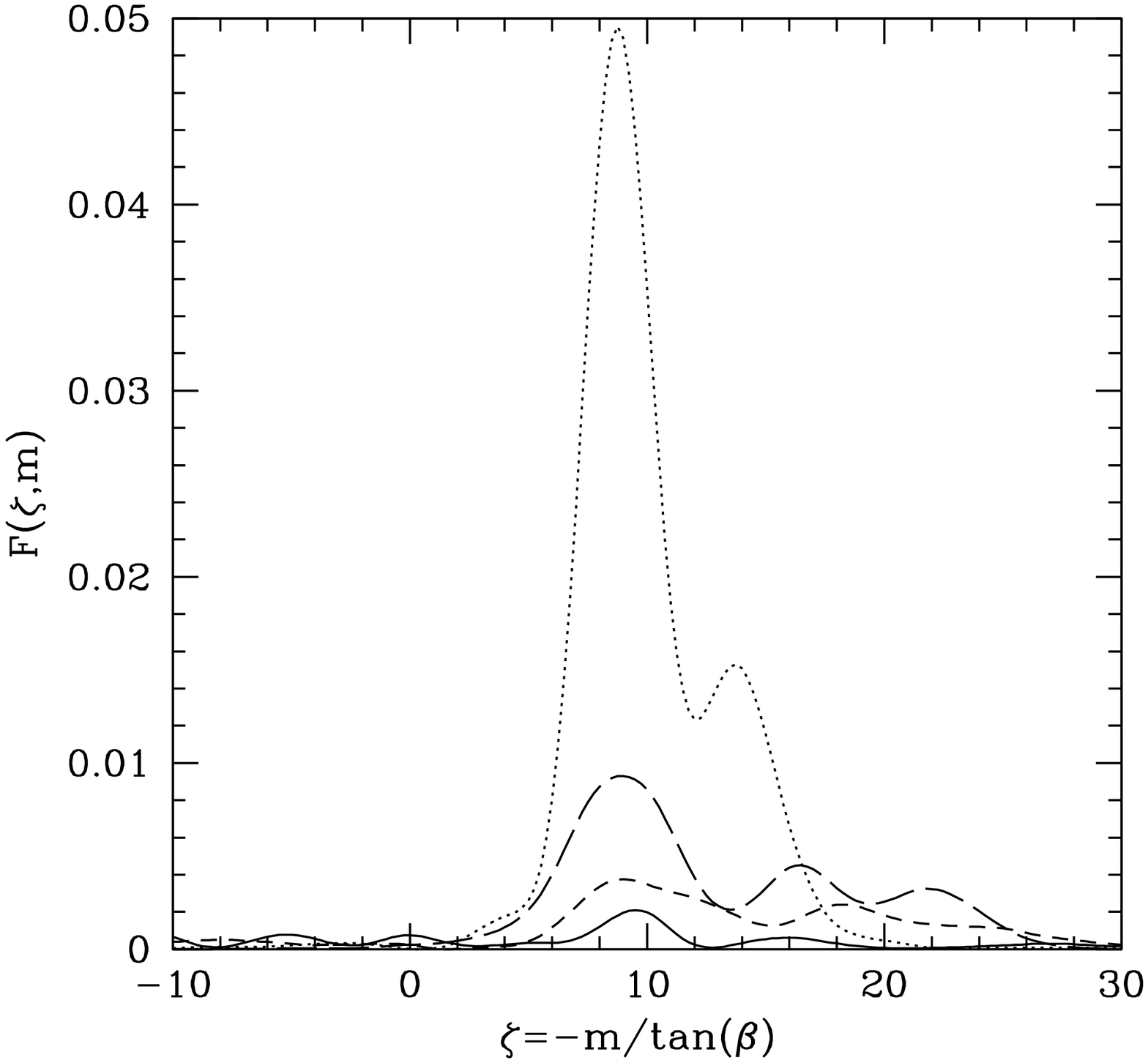}
\includegraphics[height=6.8cm]{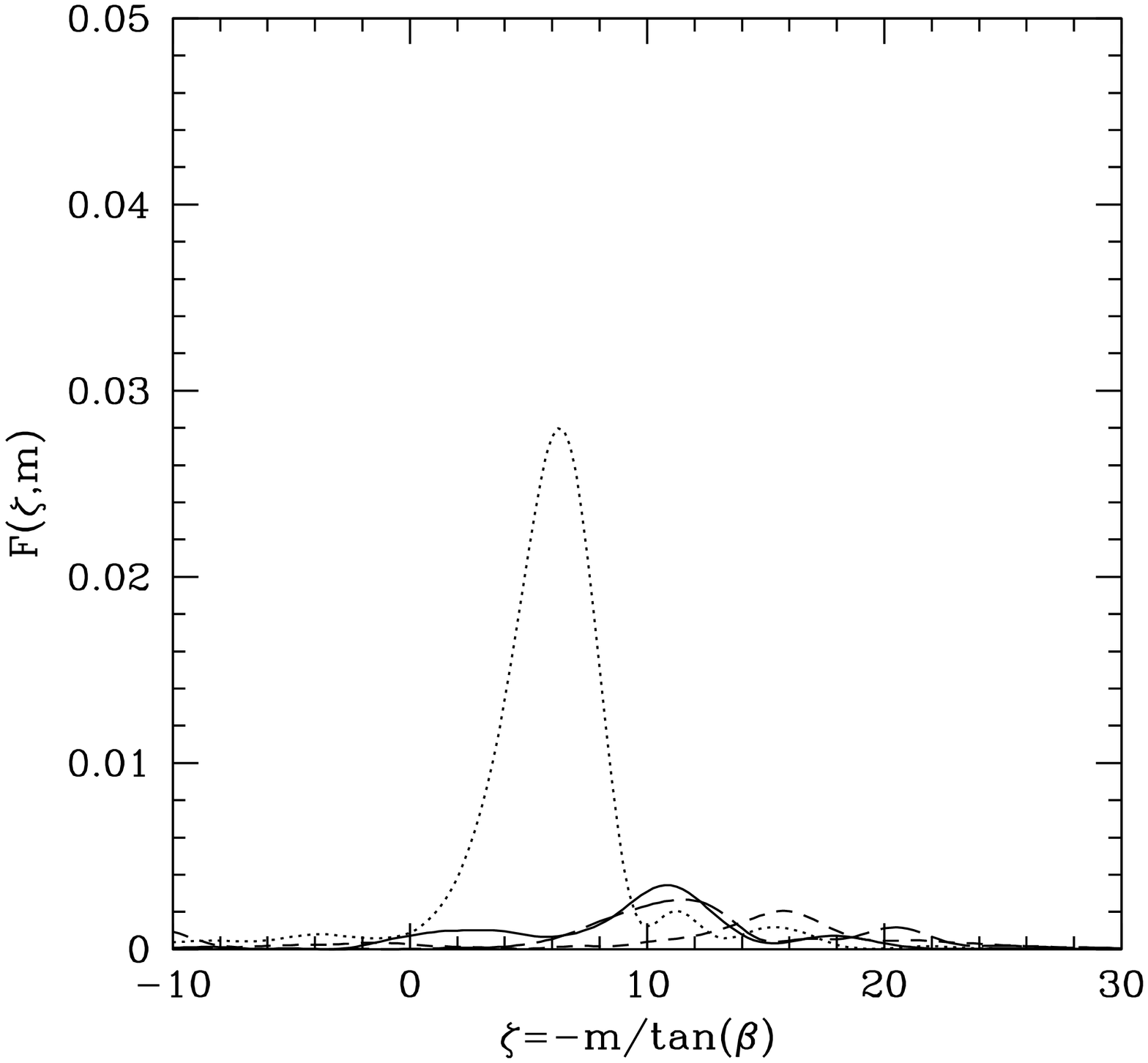}
\includegraphics[height=6.8cm]{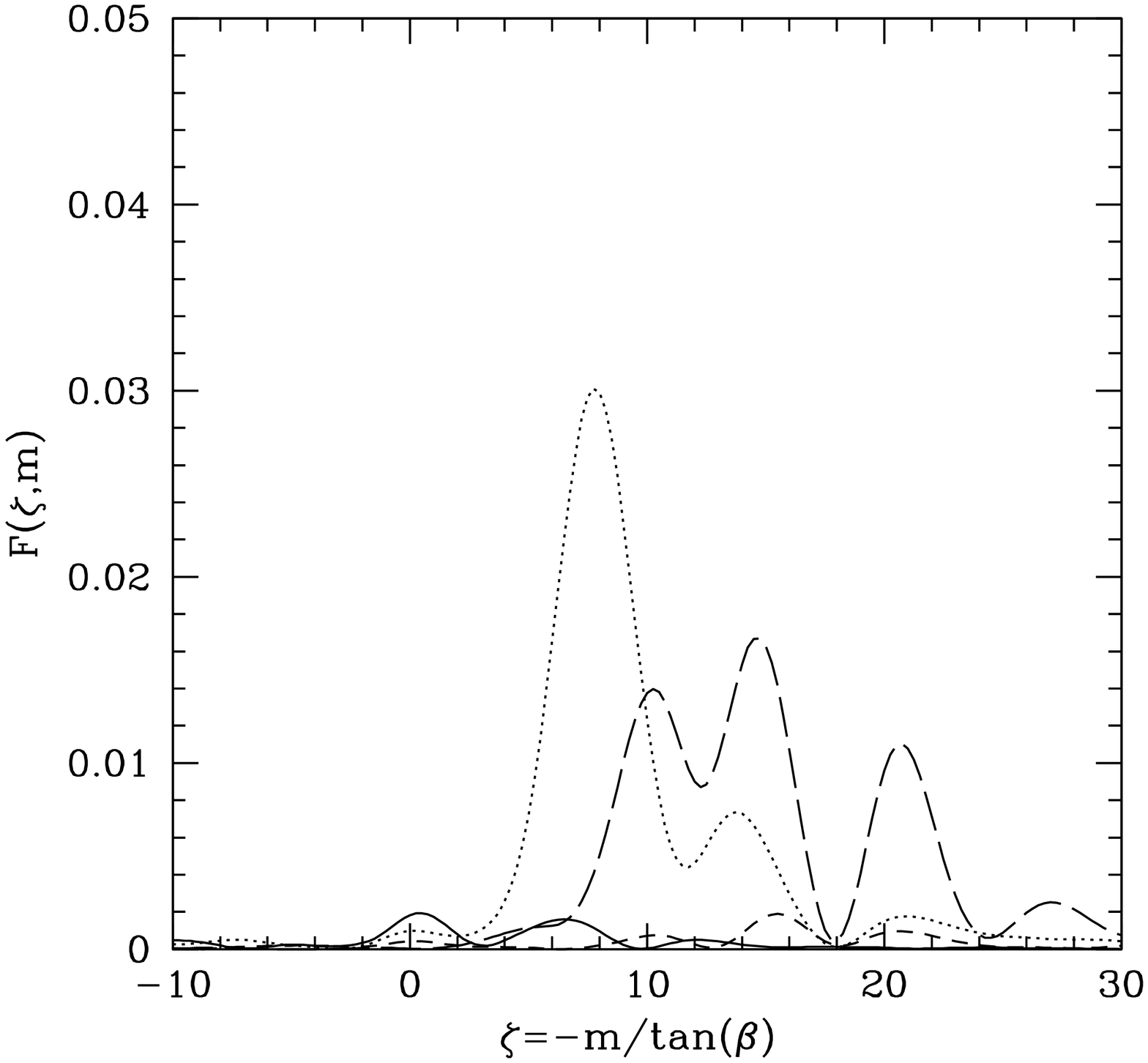}
}
\caption{The relative strengths, $F$, of the $m=1,2,3$ and 4 spiral modes (solid, dotted, short dash, long dash lines respectively)  at $t=3$~kyr (just before the discs start to fragment), against $\zeta=-m/\tan(\beta)$; $\beta$ is the pitch angle. Three graphs are plotted, (i) for the  Stamatellos et al. (2007a) diffusion approximation (top) (ii) for the Goodwin et al. (2004a) barotropic EOS (middle), and (iii) for the Bate et al. (2003) barotropic EOS (bottom). The $m=2$ is the strongest mode in all three treatments. However, the $m=4$ mode has also a significant strength for the Bate et al. barotropic EOS.}
\label{fig:fourier}
\end{figure}

The disc fragments  with all three treatments. However, the path to fragmentation and the outcome of the fragmentation is different in the three cases (Figs.~\ref{fig:images}-\ref{fig:fourier}).

%%%%%%%%%%%%%%%%%%%%%%%%%%%%%%%%%%%%%%%%%%%%%% 
%%%%%%%%%%%%%%%%%%%%%%%%%%%%%%%%%%%%%%%%%%%%%%

%%%%%%%%%%%%%%%%%%%%%%%%%%%%%%%%%%%%%%%%%%%%%
%%%%%%%%%%%%%%%%%%%%%%%%%%%%%%%%%%%%%%%%%%%%%

%%%%%%%%%%%%%%%%%%%%%%%%%%%%%%%%%%%%%%%%%%%%%%
\begin{table}
\begin{minipage}{0.45\textwidth}
\caption{The properties of the fragments formed in the disc, using (i) the Stamatellos et al. (2007a) diffusion approximation (identifier S; the colour ID refers to Fig.~\ref{fig:fragments}) (ii) the Goodwin et al. (2004a) barotropic EOS (G), and (iii) the Bate et al. (2003) barotropic EOS (B). $R_i$ (AU) is the radius of the fragment at formation (i.e. when a sink is formed), $M_f$ is the mass of the fragment at the end of the simulation, and $t_i$ the formation time. $\delta t$ is the time between the formation of the first core ($\rho\sim5\times10^{-12}$ g cm$^{-3}$) and the formation of the second core ($\rho\sim10^{-3}$ g cm$^{-3}$) for the simulation using the diffusion approximation.}
\label{tab:fragments}
\centering
\renewcommand{\footnoterule}{}  % to avoid a line before footnotes
\begin{tabular}{@{}lcccc} \hline
\noalign{\smallskip}
ID
&$R_i$ (AU)
& $M_f$ (M$_{\sun}$)
&$t_i$
& $\delta t$ (kyr) 
\\
\noalign{\smallskip}
\hline
\noalign{\smallskip}
      S1 (red, dashed) &    75  &  0.174  & 5.0 & 1.9   \\
      S2 (green) &   115   & 0.139 & 5.9 &2.8     \\
      S3 (blue) &    305 &  0.097 & 7.0 &3.4    \\
      S4 (cyan) &    140  & 0.049 & 7.3 &7.3     \\
      S5 (red) &    170  & 0.058 & 8.9 &4.9     \\
      S6 (black, dashed) &    385 &  0.037 & 9.9 &3.7      \\
      S7 (black) &    310 &  0.031 & 12.6 &7.3     \\
\noalign{\smallskip}
\hline
\noalign{\smallskip}
      G1 &    120  &  0.209 & 10.9 & - \\
      G2 &    240  &  0.135 & 10.9  & -  \\
\noalign{\smallskip}
\hline
\noalign{\smallskip}
      B1 &    180  &  0.171 & 4.1   & -  \\
      B2 &    250  &  0.092  & 4.6   & - \\
      B3 &    150 &   0.065  & 4.7   & -  \\
      B4 &    120  &  0.070  & 4.9   & -  \\
      B5 &      30  &  0.070  & 5.3   & -  \\
      B6 &     30   &  0.121 & 5.4   & -  \\
      B7 &   100   &  0.021 & 6.5   & -  \\
      B8 &       2   &  0.006 & 9.4   & -  \\
 \noalign{\smallskip}
\hline
\end{tabular}
\end{minipage}
\end{table}
%%%%%%%%%%%

%%%%%%%%%%%%%%%%%%%%%%%%%%%%%%%%%%%%%%%%%%%%%
\begin{figure}
\centering{
\includegraphics[height=8.3cm,angle=-90]{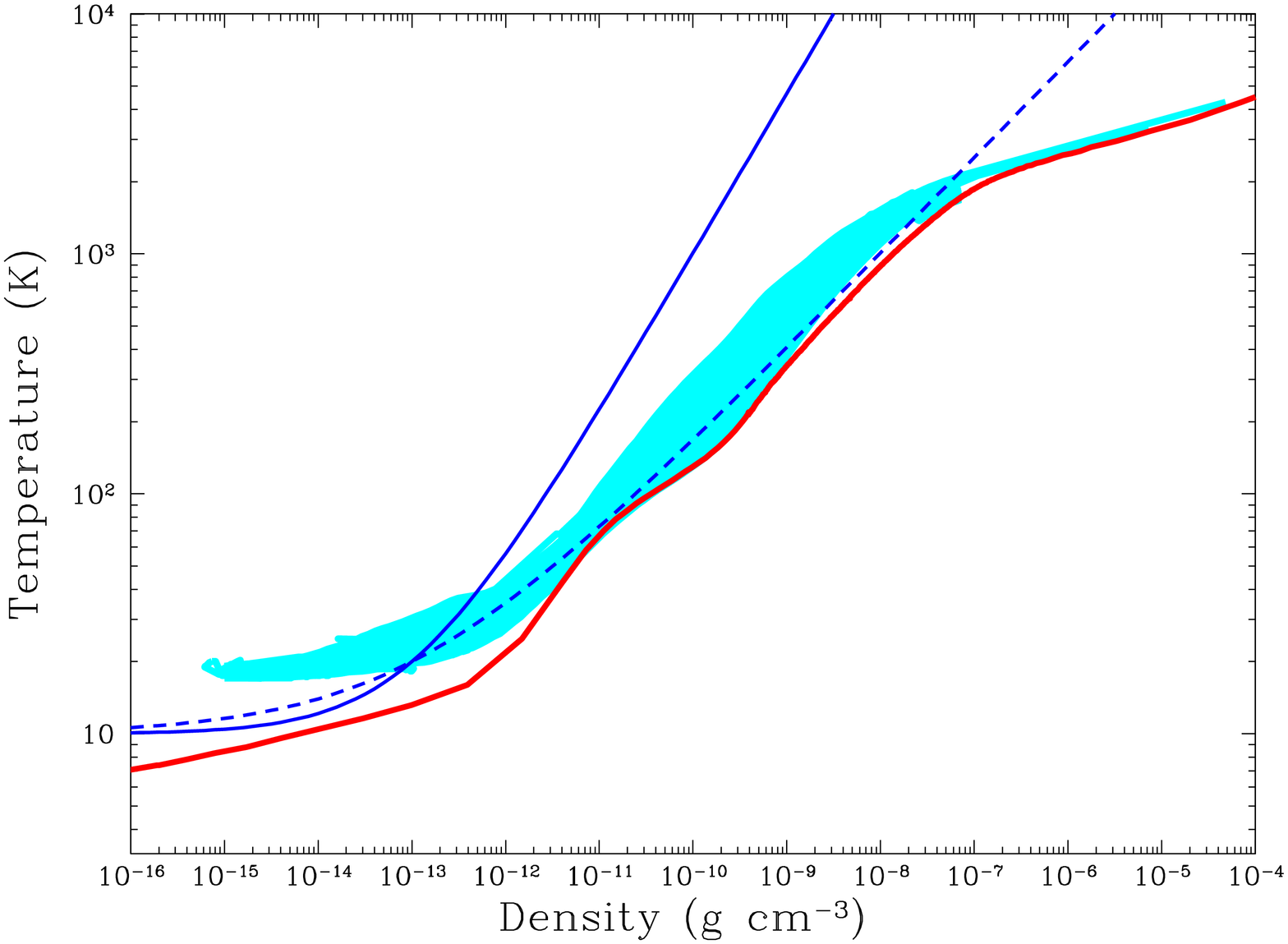}
\includegraphics[height=8.3cm,angle=-90]{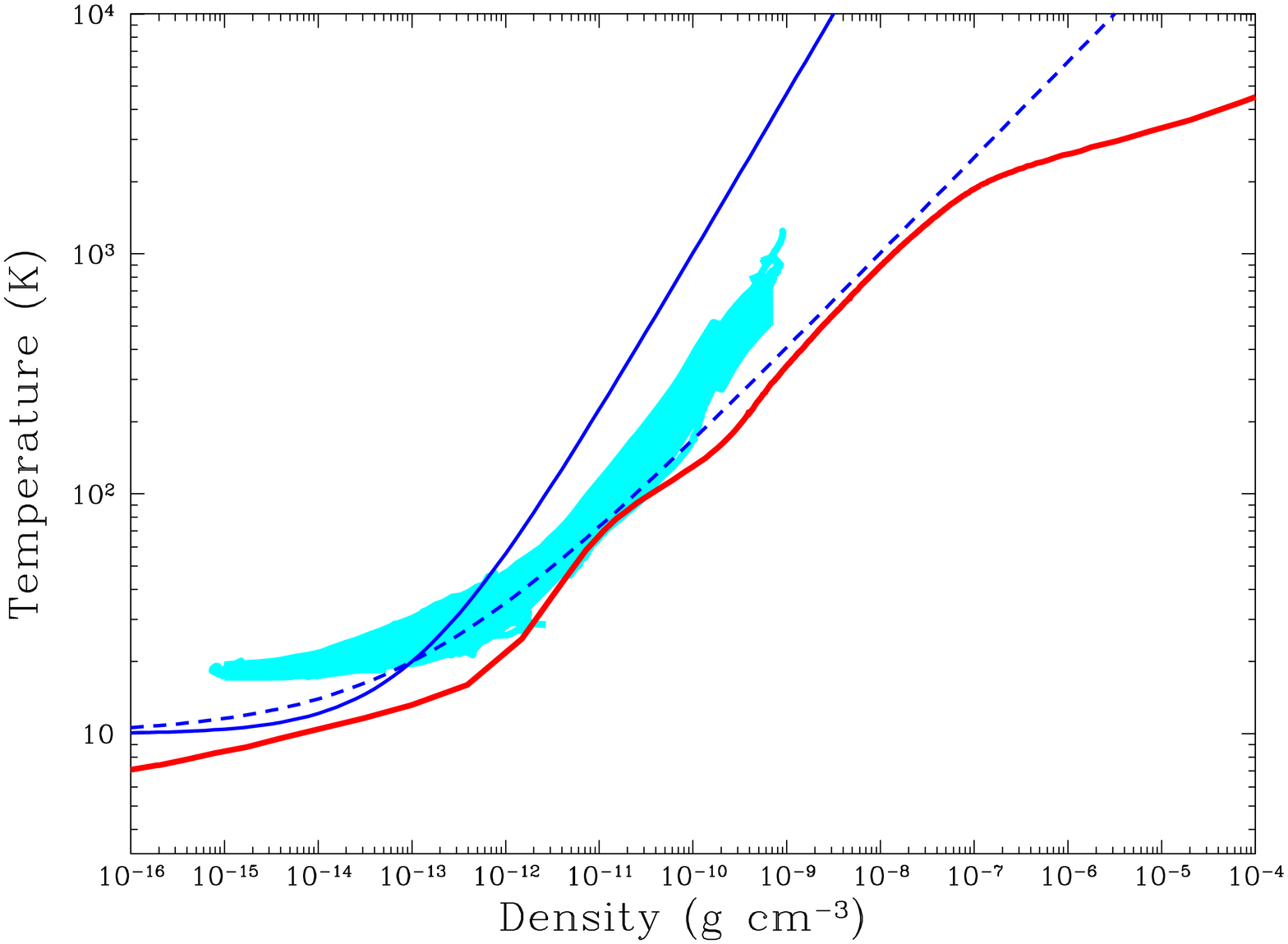}
}
\caption{The evolution of the density and the temperature at the centre (top) and at distance $r=1$~AU from the centre of each proto-fragment forming in the disc, using (i) the Stamatellos et al. (2007a) diffusion approximation (shaded cyan) (ii) the Goodwin et al. (2004a) barotropic EOS (solid blue line), and (iii) the Bate et al. (2003) barotropic EOS (dashed blue line). Proto-fragments forming farther out from the central star tend to be hotter (i.e. lie in the upper part of the shaded region) than proto-fragment forming close to the central star. The evolution of the density and temperature at the centre of a 1~M$_{\sun}$ collapsing cloud core is also plotted for reference (solid red line; Stamatellos et al. 2007a). }
\label{fig:td}
\end{figure}
%%%%%%%%%%%%%%%%%%%%%%%%%%%%%%%%%%%%%%%%%%%%%

Initially, before the disc fragments, the instabilities develop with different patterns.  To quantify these differences, we decompose the disc structure at t= 3~kyr, i.e. just before the discs fragment, into a sum of Fourier components. We use as our basis a logarithmic spiral, $R=R_0 e^{-m \phi /\zeta}$, where $m$ is the mode of the perturbation, $\phi$ is the azimuthal angle of the SPH particle, and $\zeta=-m/\tan(\beta)$ is a parameter that represents the pitch angle $\beta$ of the spiral (Sleath \& Alexander 1996; Stamatellos \& Whitworth 2008). The Fourier transform is then
\begin{eqnarray}\nonumber\label{eq:fourier}
F(\zeta,m) &= &\frac{1}{N}\sum_{j=1}^{N}{e^{-i\left(\zeta\ln[R_j]+m\phi_j\right)}},
\end{eqnarray}
where $(R_j,\phi_j)$ are the co-ordinates of particle $j$. In Fig.~\ref{fig:fourier} we plot $F(\zeta,m)$ against $\zeta$ for the $m=1,2,3$ and 4 modes, for the three simulations.  The maxima in $F(\zeta,m)$ identify the dominant pitch angles of the spirals. The most dominant mode is the $m=2$ in all three cases. However, in the case of the Bate et al. barotropic EOS there is also a strong $m=4$ mode. This results in a ring-like structure forming in the disc at $\sim 200$~AU (Fig.~\ref{fig:images}). This ring-like structure is the first to fragment, whereas in the other two cases fragmentation happens first in the inner disc region and then continues in the outer disc region.

The number of fragments produced with the harder (i.e. steeper) barotropic EOS of Goodwin et al. (2004a) is smaller  than the number of objects produced with the other two treatments (Fig.~\ref{fig:images}; Table~\ref{tab:fragments}). This is expected as with a harder EOS the matter is hotter for a given density; the increased thermal pressure suppresses fragmentation. In this simulation only 2 objects form (initially there are 3 proto-fragments but two of them coalesce before they form sinks). The two objects have relatively high masses (0.14 and 0.21 M$_{\sun}$, i.e. they are low-mass hydrogen burning stars; Fig.~\ref{fig:mrsink}). 

The other two treatments produce similar numbers of objects (7 and 8, for the diffusion approximation and the Bate et al. barotropic EOS, respectively). However there are differences in the properties of these objects (cf. Fig.~\ref{fig:mrsink}; Table~\ref{tab:fragments}). With the Stamatellos et al. (2007a) diffusion approximation, fragmentation happens first close to the central star; the more massive objects form closer to the central star in the inner disc region where there is more matter for fragments to accrete. Using the Bate et al. (2003) barotropic EOS, low-mass objects (brown dwarfs) form closer to the central star, whereas low-mass H-burning stars form in the outer disc region; the outer disc region fragments first. This is probably due to the difference in the way the instabilities grow with these treatments. Using the Bate et al. barotropic EOS a ring-like structure first forms at $\sim 200$~AU and fragmentation happens first there; the objects produced in the ring region are the most massive in the simulation. When the radiative transfer is treated with the diffusion approximation the $m=2$ mode dominates and instabilities grow faster in the inner disc region where there is more disc mass. In this case the objects that form in the inner disc region are the more massive ones.

The evolution of the density and temperature at the centre of each  proto-fragment is shown in Fig.~\ref{fig:td} (top graph). The density and temperature for the barotropic EOSs are prescribed and follow Eq.~\ref{eq:beos}. The results for the diffusion approximation are shown in cyan (for all 7 of the proto-fragments). The density and temperature at the centre of each proto-fragment are more similar to the values given by the barotropic  EOS of Bate et al. (2003). However, there is a spread due to the fact that the diffusion approximation accounts (i) for environmental factors that affect the transport of radiation around each proto-fragment, and (ii) for thermal inertia effects, which are important for fragments forming fast in the outer disc regions (see next Section). The differences are more prominent at 1~ AU from the centre of the proto-fragment  (Fig.~\ref{fig:td}, bottom graph). The temperature, for a given density, is larger when the radiative transfer is treated properly as the energy diffuses from the accretion shock around the proto-fragment, and heats its surroundings.

We thus conclude that the barotropic EOS can capture only the general way that gravitational instabilities develop in a disc; a detailed treatment of the radiative transfer results in a different disc structure, a different path to fragmentation, and hence different  properties (mass, formation radius) of the objects formed. 

%%%%%%%%%%%%%%%%%%%%%%%%%%%%%%%%%%%%%%%%%%%%%
\begin{figure}
\centering{
\includegraphics[width=6.5cm,angle=-90]{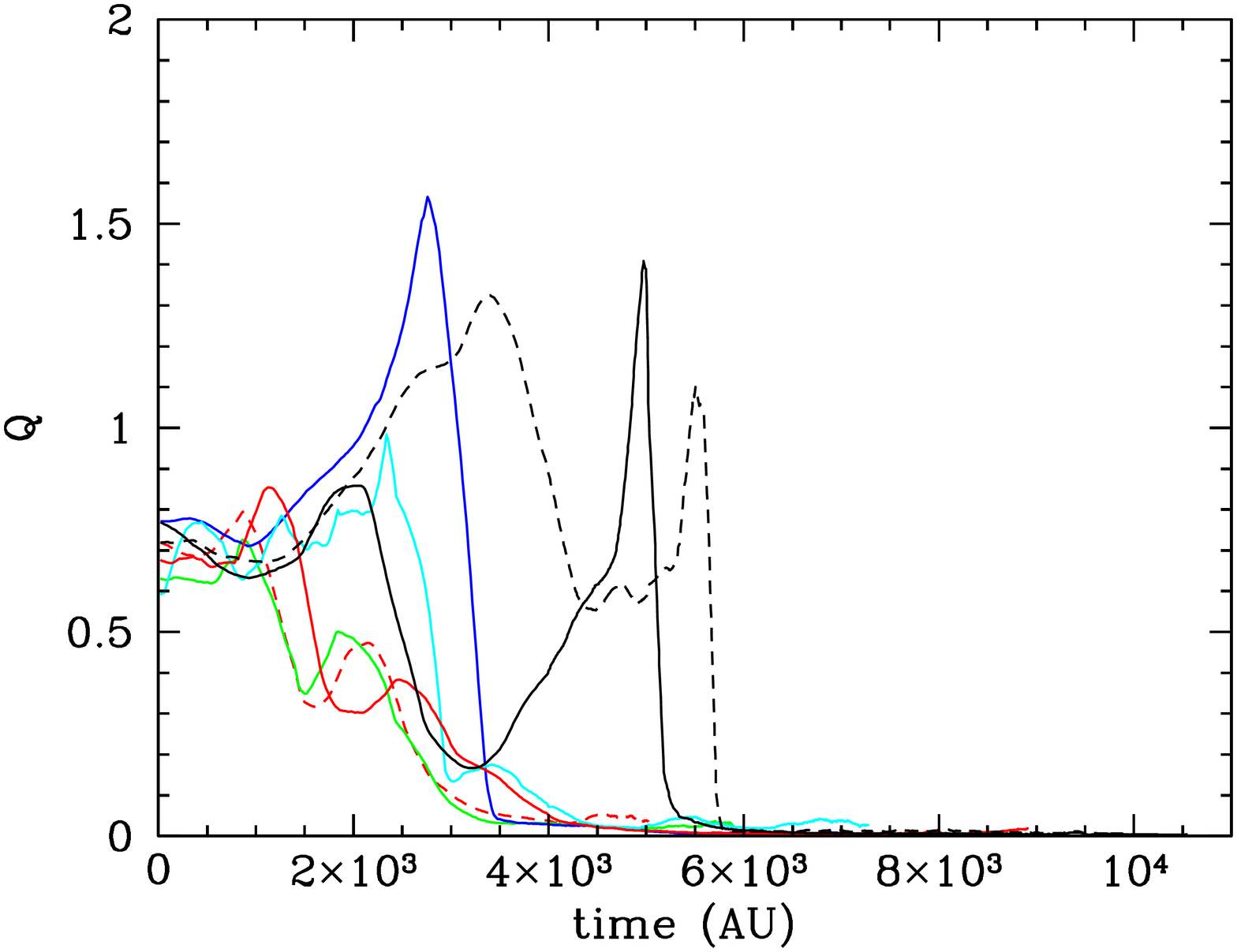}
\includegraphics[width=6.5cm,angle=-90]{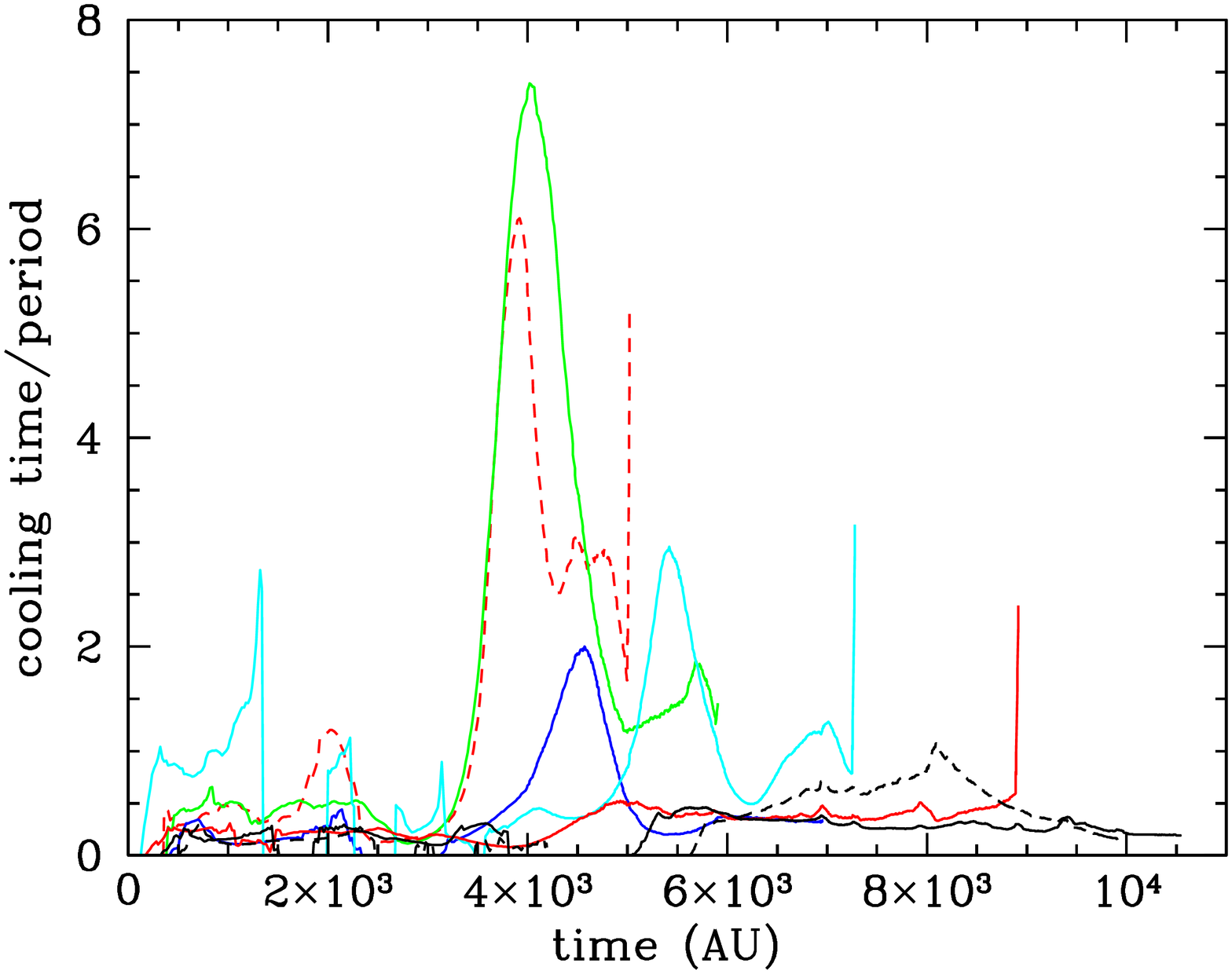}}
\caption{The Toomre parameter (top) and the cooling time in units of the local orbital period (bottom) of the proto-fragments forming in the disc.}
\label{fig:fragm.conds}
\end{figure}
%%%%%%%%%%%%%%%%%%%%%%%%%%%%%%%%%%%%%%%%%%%%%

%%%%%%%%%%%%%%%%%%%%%%%%%%%%%%%%%%%%%%%%%%%%%
\begin{figure}
\centering{
\includegraphics[width=\columnwidth]{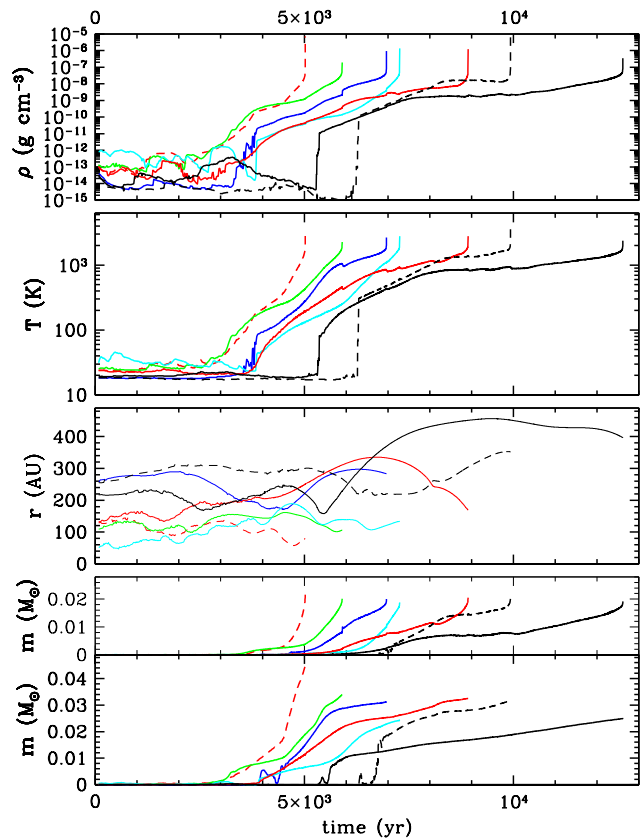}}
\caption{The evolution of the properties of the proto-fragments formed in the disc using the Stamatellos et al. (2007a) diffusion approximation. From top to bottom for each proto-fragment are plotted (until each one is replaced by a sink at $10^{-3}$ g cm$^{-3}$)  (i) the density at the centre of the proto-fragment, (ii) the temperature at the centre of the proto-fragment, (iii) the distance of the proto-fragment  from the central star, (iv) the mass within 1~AU from the centre of the proto-fragment, and (v) the mass within  5~AU from the centre of the proto-fragment.}
\label{fig:fragments}
\end{figure}
%%%%%%%%%%%%%%%%%%%%%%%%%%%%%%%%%%%%%%%%%%%%%

\section{The evolution of proto-fragments forming at different disc radii}

We now concentrate on the characteristics of the evolution of the proto-fragments that form in the disc when the radiative transfer is treated with the  diffusion approximation of Stamatellos et al. (2007a).

The Toomre ($Q<1$) and the Gammie ($t_{_{\rm COOL}}/t_{_{\rm ORB}}<2$) conditions for fragmentation are met for all proto-fragments (Fig.~\ref{fig:fragm.conds}), at least for a large period during their evolution. The cooling time for some of the proto-fragments that condense out closer to the central star is larger than the Gammie critical value for a period of up to $\sim 1$~kyr but this does not seem to affect the long-term evolution of these proto-fragments).

The proto-fragments forming in the disc go through the same phases as a collapsing 1 M$_{\sun}$  core (e.g. Larson 1969; Masunaga \& Inutsuka 2000; Stamatellos et al. 2007a). During the initial stage of their formation the proto-fragments stay almost isothermal as they are optically thin and the gravitational energy that is converted to thermal energy can be readily radiated away. The temperature of the proto-fragments at this stage is $\sim 10$~K, for the ones forming away from the central star, and up to $\sim20-30$~K, for the ones that form close to central star (due to heating from the star) (Fig.~\ref{fig:fragments}, T). The proto-fragments get denser, become optically thick, and start heating up. The increased thermal pressure slows down the collapse, and quite often the proto-fragments experience an adiabatic bounce  (Fig.~\ref{fig:fragments}, $\rho$). When a proto-fragment cannot cool fast enough, i.e. close to the central star, this bounce is followed by the shearing apart of the proto-fragment, due to gravitational torques. Proto-fragments that form farther away from the central star ($\stackrel{>}{_\sim} 60-70$~AU) survive and as they accrete material from the disc, they start to contract once again (and maybe bounce again) until  the first hydrostatic core forms ($\rho\sim 5\times10^{-12}$ g~cm$^{-3}$). This core then contracts quasi-statically on a Kelvin-Helmholtz timescale. The first core grows gradually in mass, and its density and  temperature increase slowly. When the temperature at its centre reaches 2000~K, the second collapse starts, as the energy delivered by the compression does not heat the core further, but instead goes into the dissociation of molecular hydrogen. The second collapse leads to the formation of the second core (i.e. the protostar or proto- brown dwarf).

The above general characteristics are the same for all proto-fragments forming in the disc but there are notable differences in the way that proto-fragments condense out in the disc at different distances from the central star. Generally speaking, proto-fragments closer to the central star start forming  earlier than proto-fragments farther away, as the gravitational instabilities develop faster in the inner disc region. They grow in mass gradually and take about  $\sim 3-4$~kyr,  to reach the stage of the first core.  On the other  hand, proto-fragments that form farther away from the central star, they start forming at later times, but they reach the first core stage by growing in mass rather abruptly ($\sim 20-100$~yr) due to spiral arms interacting with each other, and interactions and/or mergers with other proto-fragments (see the sharp increase of density, temperature and mass within 5~AU for the last two proto-fragments forming in the disc at $t\sim 40$~kyr and at $t\sim 50$~kyr, respectively; Fig~\ref{fig:fragments}).

The first core of a proto-fragment  grows in mass gradually through accretion of material from the disc, and sometimes episodically with help from special accretion events, such as  interactions and/or mergers with other proto-fragments in the disc  or a passage from a spiral arm. Proto-fragments closer to the central star evolve  faster from the first to the second core  than proto-fragments forming in the outer disc region (Table~\ref{tab:fragments}). The main reason for this is that  there is more mass in the inner disc region for the proto-fragments to accrete and grow. There is an inverse correlation between the mass available within 5~AU from the centre of a proto-fragment (Fig.~\ref{fig:fragments}, bottom) and the time that the proto-fragment takes to evolve from the first core to the second core  (Table~\ref{tab:fragments}). In the absence of a pool of material available for accretion it is possible that these proto-fragments might evolve on a much longer timescale (on the order of a few tens of kyr). Another reason for the longer time needed for proto-fragments forming farther out in the disc to evolve to the second core is that they tend to be hotter (at a specific density) than proto-fragments forming closer to the central star; the temperature vs density relation for the former tends to be on the upper part of the cyan shaded region in Fig.~\ref{fig:td}, whereas for the latter tends to be on the lower part of the shaded region. This is due to the timescale of formation of the first core. As mentioned previously, proto-fragments in the inner disc region accumulate the mass of the first core over a period of $\sim 3-4$~kyr, whereas proto-fragments in the outer disc region do that much faster, within $\sim 20-100$~yr. Thus, the former radiate away the energy provided by the collapse gradually as the first core forms (the radiative cooling timescale is on the order of 1-10~kyr), whereas the latter have to radiative this energy away after the first core forms. This is a thermal inertia effect, and thus cannot be captured using a barotropic EOS.

The initial masses of the first cores  (i.e. within $\sim 1$~AU) are similar for all the proto-fragments forming in the disc ($\sim 20$~M$_{\rm J}$). However proto-fragments closer to the central star will accrete more mass from the disc and become the higher mass objects formed in the disc (high-mass brown dwarfs or low-mass H-burning stars; Stamatellos \& Whitworth 2009a,b).

%%%%%%%%%%%%%%%%%%%%%%%%%%%%%%%%%%%%%%%%%%%%%
\begin{figure}
\centering{
\includegraphics[width=6.5cm,angle=-90]{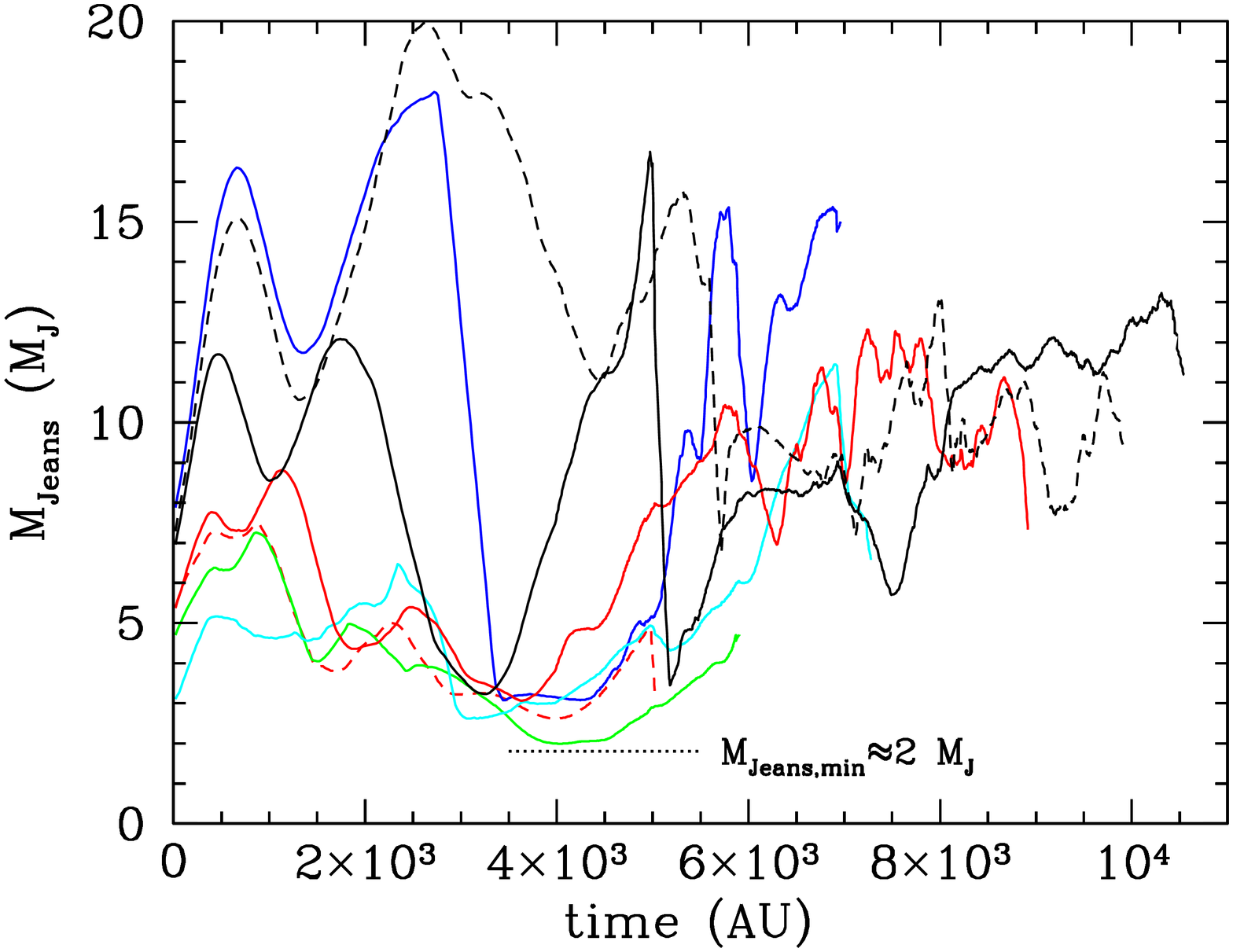}
\includegraphics[width=6.5cm,angle=-90]{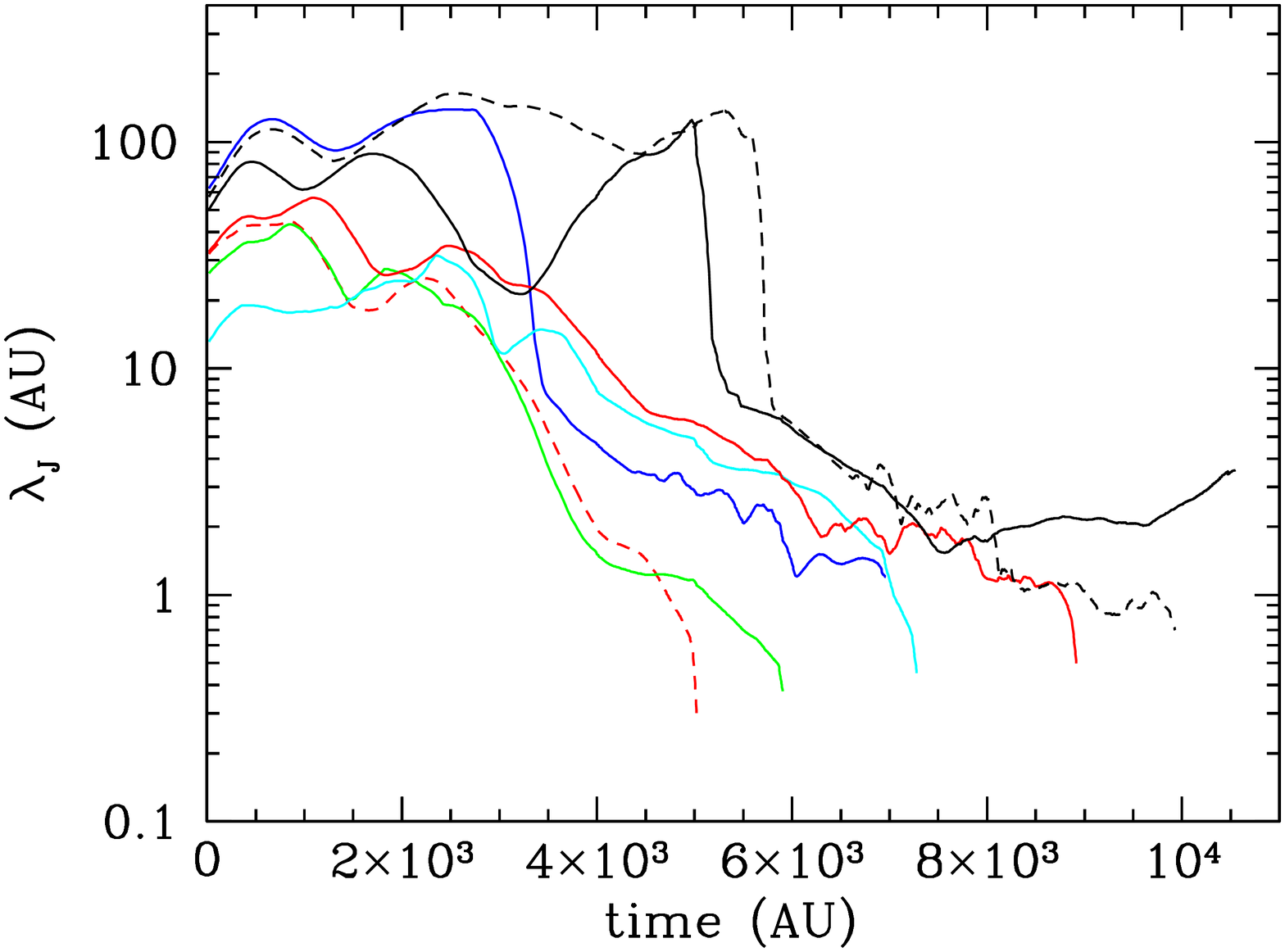}}
\caption{The Jeans mass (in Jupiter's masses; top) and the Jeans wavelength (in AU; bottom) of the proto-fragments forming in the disc. The minimum Jeans mass in our simulation is $M_{_{\rm JEANS, MIN}}\approx 2 M_{\rm J}$). This mass is adequately resolved in the simulation (corresponds to $\sim8\times N_{_{\rm NEIGH}}$).}\label{fig:ml.jeans}
\end{figure}
%%%%%%%%%%%%%%%%%%%%%%%%%%%%%%%%%%%%%%%%%%%%%

%%%%%%%%%%%%%%%%%%%%%%%%%%%%%%%%%%%%%%%%%%%%%
\begin{figure}
\centering{
\includegraphics[width=6.5cm,angle=-90]{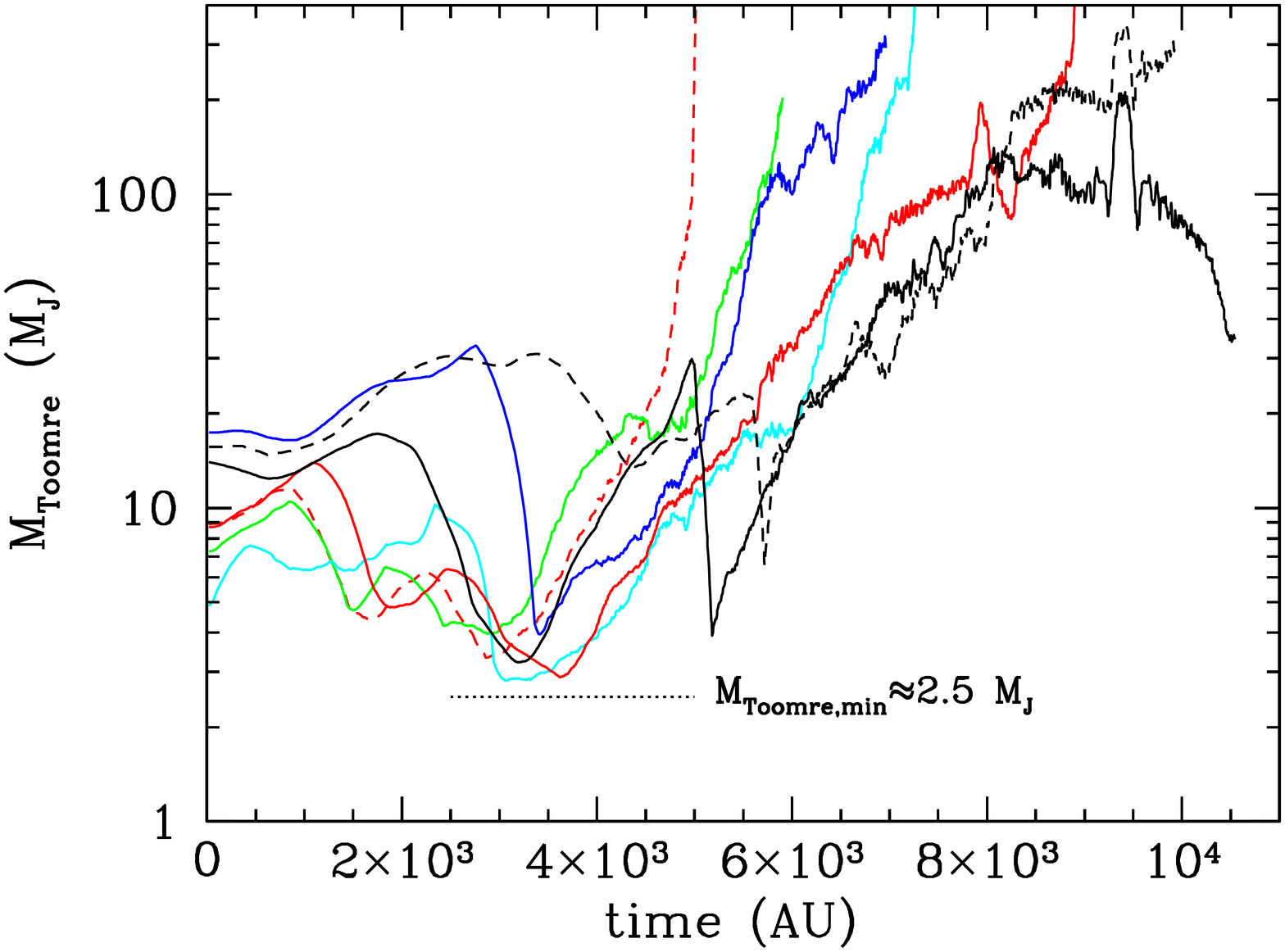}
\includegraphics[width=6.5cm,angle=-90]{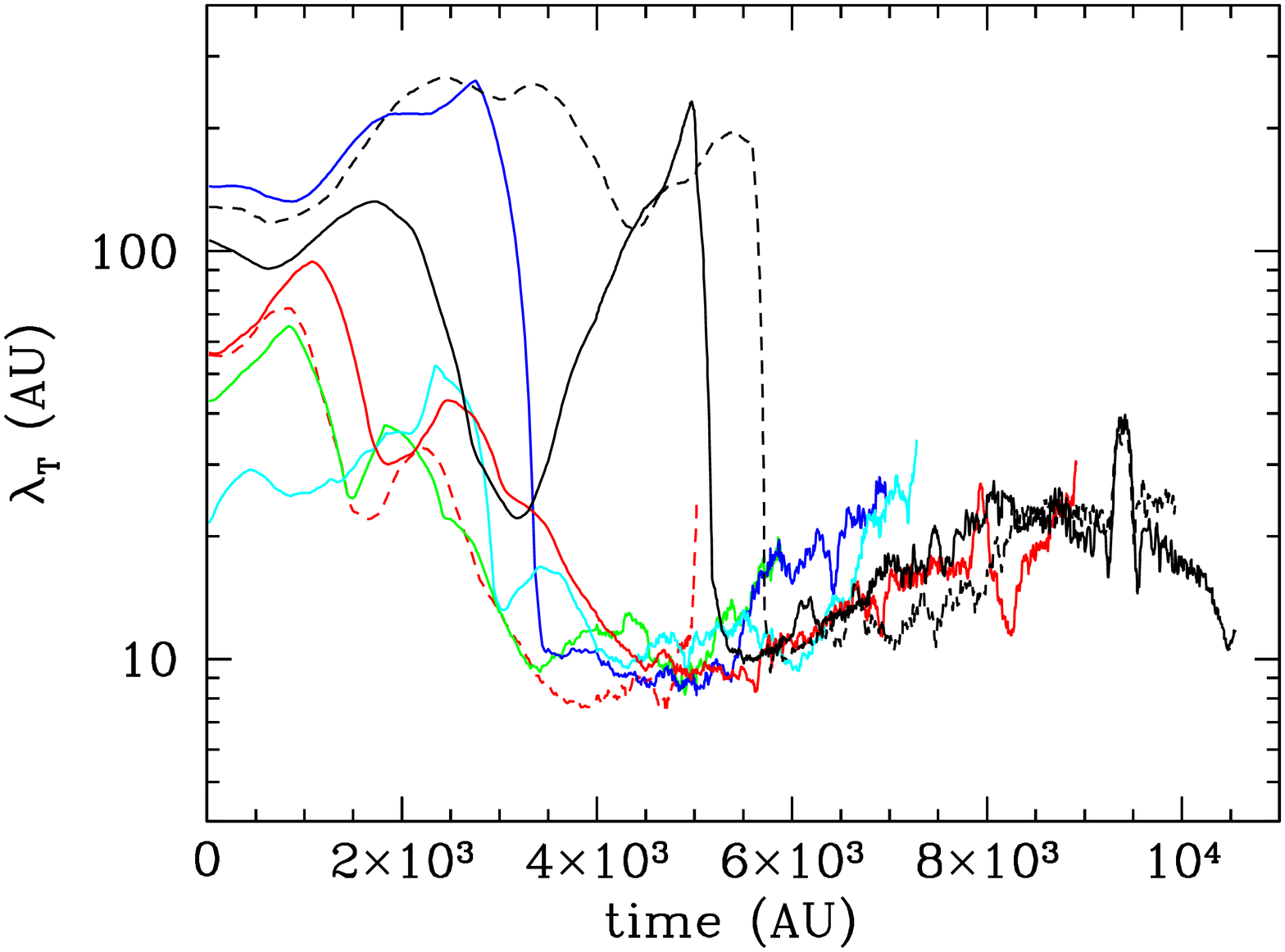}}
\caption{The Toomre mass (in Jupiter's masses; top) and the Toomre wavelength (in AU; bottom) of the proto-fragments forming in the disc. The minimum Toomre mass in our simulation is $M_{_{\rm TOOMRE, MIN}}\approx 2.5 M_{\rm J}$). This mass is adequately resolved in the simulation (corresponds to $\sim10\times N_{_{\rm NEIGH}}$). }
\label{fig:ml.toomre}
\end{figure}
%%%%%%%%%%%%%%%%%%%%%%%%%%%%%%%%%%%%%%%%%%%%%

\section{Numerical requirements of disc simulations}

 We use 150,000 particles to represent the disc, which means that the minimum resolvable mass (corresponding to the number of neighbours used, i.e. 50 SPH particles) is $\approx 0.25\,{\rm M}_{_{\rm J}}$. We have also performed simulations using 250,000 and 400,000 particles (Stamatellos \& Whitworth 2009a). In these simulations the growth of gravitational instabilities, and the properties of the proto-fragments formed as a result of these instabilities, follows the same patterns as in the simulation with lower resolution.  The final outcome is different in detail, but this is to expected as (i) these simulations have different seed noise (due to random positioning of the particles), and (ii) the gravitational instabilities develop in a chaotic, non-linear way. The simulations appear to be converged, in a statistical sense.

Nelson (2006) set a number of numerical requirements for disc simulations. These requirements were devised in the context of the specific simulations performed by Nelson (2006), hence they have limitations. Nevertheless, they provide useful guidelines for the resolution requirements of disc simulations. In the next subsections we discuss how our simulation satisfies these requirements. 

\subsection{Resolving the Jeans mass and the Toomre mass}

It has been shown that in order to avoid numerical (artificial) fragmentation the spatial resolution of the simulation has to be smaller that the local Jeans wavelength 
\begin{equation}
\lambda_J=\left(\frac{\pi c_s^2}{G\rho}\right)^{1/2}
\end{equation}
by a factor of a few ($\sim 4$; Truelove et al. 1997). For SPH simulations this means that the local Jeans mass 
\begin{equation}
M_J=\frac{4\pi^{5/2}}{24}\frac{c_s^3}{(G^3\rho)^{1/2}}
\end{equation}
must be resolved by $2\times N_{_{\rm NEIGH}}$ particles (Bate \& Burkert 1997).
The Jeans mass and the Jeans wavelength of the proto-fragments forming in the disc are shown in Fig.~\ref{fig:ml.jeans}. The minimum Jeans mass in our simulation is $M_{_{\rm JEANS,MIN}}\approx 2 M_{\rm J}$. Thus, according to  the Bate \& Burkert (1997) condition, this mass is adequately resolved in the simulation as it corresponds to $\sim8\times N_{_{\rm NEIGH}}$ ($N_{_{\rm NEIGH}}=50$).

Nelson (2006) argues that in disc simulations it is more appropriate to be able to resolve the Toomre length
\begin{equation}
\lambda_T=\frac{2 c_s^2}{G\Sigma}
\end{equation}
and the corresponding Toomre mass
\begin{equation}
M_T=\frac{\pi c_s^4}{G^2\Sigma}\,.
\end{equation}
By performing numerical experiments Nelson (2006) concludes that the Toomre mass must be resolved by a minimum of $6\times N_{_{\rm NEIGH}}$. The Toomre mass  and the Toomre wavelength  of the proto-fragments forming in the disc are shown in Fig.~\ref{fig:ml.toomre}. The minimum Toomre mass in our simulation is $M_{_{\rm TOOMRE, MIN}}\approx 2.5 M_{\rm J}$). This mass is adequately resolved as it corresponds to $\sim10\times N_{_{\rm NEIGH}}$. 

\subsection{Variable gravitational softening}

We use a variable gravitational softening that is set equal to the smoothing length, i.e. the length scale over which hydrodynamic quantities are smoothed. Thus, according to Whitworth (1998) and Nelson (2006) we avoid artificial suppression or enhancement of structure.

\subsection{Resolving the vertical disc structure}

Nelson (2006) suggests that in order to avoid underestimating the disc midplane density the vertical disc structure must be resolved by at least $\sim 4$ smoothing lengths per scale height. We note however that this criterion has been devised for isothermal, equilibrium discs and it is uncertain (i) how it relates to self-gravitating discs, and (ii) how it affects fragmentation.

\begin{figure}
\centering{
\includegraphics[width=6.5cm,angle=-90]{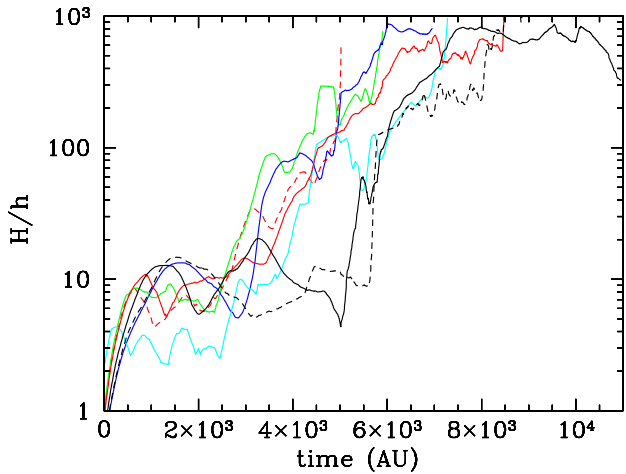}}
\caption{The ratio of disc scale height $H$ to local smoothing length $h$ for each proto-fragment forming in the disc. The disc scale height is generally resolved with at least $\sim 5$ smoothing lengths (apart from the fragment forming closer to the central star; cyan line), satisfying the Nelson (2006) condition.}
\label{fig:res.column}
\end{figure}

In Fig.~\ref{fig:res.column} we plot the ratio of disc scale height $H$ to local smoothing length $h$  for each proto-fragment forming in the disc. The scale height is calculated by assuming that the disc extends vertically up to $3H$. The disc scale height is generally resolved with at least $\sim 5$ smoothing lengths, satisfying the Nelson (2006) condition. Hence, the disc is resolved adequately despite the fact that the number of particles used is less that the number suggested by Nelson (2006). The reason for this is that we simulate the outer disc region where the disc thickness is relatively large; therefore a smaller number of particles is needed to resolve the vertical disc structure. 

Nelson (2006) also argues that in simulations with radiative transfer the disc photosphere must be also resolved properly in order to accurate calculate the energy radiated away. The number of particles that used in our simulation are not adequate to resolve the disc photosphere but tests have shown that the vertical disc temperature profile is properly captured with our radiative transfer method (see Stamatellos \& Whitworth 2008). More specifically we have shown that the disc temperature profile in a similar disc simulation fits well the temperature  profile that is analytically calculated by Hubeny (1990).

%%%%%%%%%%%%%%%%%%%%%%%%%%%%%%%%%%%%%%%%%%%%%%
%%%%%%%%%%%%%%%%%%%%%%%%%%%%%%%%%%%%%%%%%%%%%

\section{Discussion}

We have examined the role of thermodynamics in the properties of low-mass objects formed by fragmentation of massive, extended discs around Sun-like stars. We have focused on the regime of gravitationally unstable discs that can cool fast enough, i.e. discs that fragment. Different treatments of radiative transfer result not only in a different number of objects formed in the disc but also in objects with different characteristics, i.e. different mass and formation radius distributions. It is also expected that different radiative transfer treatments may determine whether a disc can fragment or not, but in the current paper we have not investigated this.

The disc thermodynamics were treated with 3 different prescriptions (i) with the diffusion approximation of Stamatellos et al. (2007a), (ii) with the barotropic EOS of Goodwin et al. (2004a), and (iii) with the barotropic EOS of Bate et al. (2003).  The barotropic approximations capture the general evolution of the density and temperature at the centre of each proto-fragment  (cf. Fig.~\ref{fig:td}) but  (i) they do not make any adjustments for the particular circumstances of different proto-fragments forming at different locations in the disc, and (ii) they do not take into account thermal inertia effects  (Boss et al. 2000) that are important for fast-forming proto-fragments in the outer disc region. Hence, they cannot reproduce the spread of the evolution of the density and temperature for different proto-fragments and for different distances from the centre of each proto-fragment (cf. Fig.~\ref{fig:td}). As a result, the properties of objects formed in the disc when a barotropic EOS is used are different. This is important not only for disc simulations but also for  simulations of collapsing clouds, as in these simulations frequently stars form with discs that later become unstable and fragment (e.g. Attwood et al. 2009). 

The importance of thermodynamics and radiative feedback in simulations of star formation has already been pointed out by the works of Boss et al. (2000), Attwood et al. (2009), Bate et al. (2009a,b), and Offner et al. (2009). These authors compare  simulations of star formation in collapsing  clouds using barotropic EOSs and using different approximations for the radiative transfer,  and find that indeed the treatment of radiative transfer is critical for the properties of the objects produced in the simulations. 

Boss et al. (2000) study the collapse of a rotating cloud core with a small $m=2$ perturbation and find that using the Eddington approximation for the radiative transfer the core fragments into a binary, whereas using  a barotropic EOS the collapse results in the formation of a single star.

Bate et al. (2009a,b) find that using the diffusion approximation fewer objects form, whereas Attwood et al. (2009) find the opposite; more objects form when switching from a barotropic EOS to the diffusion approximation. However, these authors compare the results obtained using the diffusion approximation with the results obtained using different barotropic EOSs; Attwood et al. (2009) use the harder barotropic EOS of Goodwin et al (2004a). As we find in the current paper, it is expected that the simulations of Attwood et al. (2009) (with the barotropic EOS of Goodwin et al. 2004a) should produce fewer objects than the simulations of Bate et al. (2009a,b) (also with a barotropic EOS). Hence, the results for Attwood et al. (2009) and Bate et al. (2009a,b) are probably consistent.

Attwood et al. (2009) and  Bate et al. (2009a,b) include radiative feedback from accretion onto young protostars but only for accretion onto the sink radius ($\sim 0.5-5$~AU). Therefore,  the bulk of radiative feedback from accretion of material onto the protostars is ignored. Offner et al. (2009) use a sub-grid protostellar model to account for this feedback and additional feedback from nuclear burning in the protostar. They find that radiative feedback tends to suppress fragmentation; fewer objects formed when compared with simulations using a barotropic EOS. Matzner \& Levin (2005) and  Whitworth \& Stamatellos (2006) estimate that for Sun-like stars radiative feedback suppresses fragmentation only close to the central star (up to $\sim 100$~AU), so radiative feedback may only be critical for rapidly-accreting central stars (Krumholz 2006).

We have also examined the difference in the way proto-fragments condense out in the disc at different distances from the central star (Fig.~\ref{fig:fragments}).  We find that proto-fragments forming closer to the central star tend to form earlier and evolve faster from the first core to the second core than proto-fragments forming in the outer disc region. The former have a large pool of material in the inner disc region that they can accrete from and grow in mass. The latter accrete more slowly and they are hotter (cf. Fig.~\ref{fig:td}) because they form fast in an abrupt event (e.g. an interaction or merger with other proto-fragments, or  spiral arm interactions in the disc).

We conclude that disc thermodynamics plays a critical role in disc fragmentation affecting not only the number of objects produced in a fragmenting disc but also their properties, i.e. their mass, radius of formation, and formation time. 

\section*{Acknowledgements}
We thank the referee for his comments that initiated the investigation in the resolution of our simulations. We also thank A. Nelson for providing clarifications about the numerical requirements for disc simulations that are discussed in Nelson (2006). Simulations were performed using the Cardiff HPC Cluster {\sc Merlin}. Colour plots were produced using {\sc splash} (Price 2007).  We acknowledge support from the STFC grant PP/E000967/1.

\label{lastpage}
\end{document}